\documentclass[aps,prl,reprint,amsmath,amssymb,superscriptaddress,nofootinbib,noeprint,nobibnotes]{revtex4-2}
\pdfoutput=1
\usepackage{graphicx}% Include figure files
\usepackage{bm}% bold math
\usepackage{mathtools}
\usepackage{float}
\usepackage{color}
\usepackage{graphicx}
\usepackage{graphicx}
\usepackage{braket}
\usepackage{times}
\makeatletter
\def\maketitle{
\@author@finish
\title@column\titleblock@produce
\suppressfloats[t]}

\begin{document}
\title{Observation of Anomalous Decay of a Polarized Three-Component Fermi Gas}

\author{Grant L. Schumacher}
\email[Corresponding author. E-mail: ]{grant.schumacher@yale.edu}
\affiliation{Department of Physics, Yale University, New Haven, Connecticut 06520, USA}
\author{Jere T. Mäkinen}
\affiliation{Department of Physics, Yale University, New Haven, Connecticut 06520, USA}
\affiliation{Yale Quantum Institute, Yale University, New Haven, Connecticut 06520, USA}
\author{Yunpeng Ji}
\affiliation{Department of Physics, Yale University, New Haven, Connecticut 06520, USA}
\author{Gabriel G. T. Assumpção}
\affiliation{Department of Physics, Yale University, New Haven, Connecticut 06520, USA}
\author{Jianyi Chen}
\affiliation{Department of Physics, Yale University, New Haven, Connecticut 06520, USA}
\author{Songtao Huang}
\affiliation{Department of Physics, Yale University, New Haven, Connecticut 06520, USA}
\author{Franklin J. Vivanco}
\affiliation{Department of Physics, Yale University, New Haven, Connecticut 06520, USA}
\author{Nir Navon}
\affiliation{Department of Physics, Yale University, New Haven, Connecticut 06520, USA}
\affiliation{Yale Quantum Institute, Yale University, New Haven, Connecticut 06520, USA}

\begin{abstract}
Systems of fermions with multiple internal states, such as quarks in quantum chromodynamics and nucleons in nuclear matter, are at the heart of some of the most complex quantum many-body problems. The stability of such many-body multi-component systems is crucial to understanding, for instance, baryon formation and the structure of nuclei, but these fermionic problems are typically very challenging to tackle theoretically. Versatile experimental platforms on which to study analogous problems are thus sought after. Here, we report the creation of a uniform gas of three-component fermions.  We characterize the decay of this system across a range of interaction strengths and observe nontrivial competition between two- and three-body loss processes.  We observe anomalous decay of the polarized (\emph{i.e.} spin-population imbalanced) gas, in which the loss rates of each component unexpectedly differ. We introduce a generalized three-body rate equation which captures the decay dynamics, but the underlying microscopic mechanism is unknown.
\end{abstract}

\maketitle

Characterizing if - and into what - a system decays has often been a pathway to discovering new quantum phenomena.  Such stability problems are ubiquitous in many-body physics, ranging from the onset of turbulence in quantum liquids~\cite{vinen2002quantum} to dissipation processes in Josephson junctions~\cite{pop2014coherent}, to the stability of nuclei~\cite{thoennessen2004reaching}.
Ultracold atomic systems have been a fertile ground for such studies: their (in)stability has revealed a wide range of interesting and sometimes unexpected phenomena, such as Efimov three-body physics~\cite{naidon2017Efimov}, quantum-fluctuation stabilization against many-body collapse~\cite{bottcher2020new}, 
and prethermalization of metastable phases~\cite{ueda2020quantum}. In particular, there has been renewed appreciation that the stability of quantum gases against inelastic-collision losses~\cite{kagan1985effect,burt1997coherence} (or lack thereof) is a pristine probe of many-body quantum correlations~\cite{laurent2017connecting,eigen2017universal,he2020universal,werner2022three}.

Strongly interacting fermions with contact interactions embody a stability success story, especially the case of two-component (`spin-1/2') fermions. The observation of unexpected shifts of loss features with respect to scattering - Feshbach - resonances~\cite{dieckmann2002decay,o2002measurement,bourdel2003measurement,regal2003creation,jochim2003pure} led to the understanding that surprisingly stable pairs were forming near those resonances, protected by a Pauli blocking effect. This discovery unlocked the field of the BEC-BCS crossover~\cite{randeria2012bcs}. 

Three-component (`spin-1') fermions give access to even richer physics, such as quantum chromodynamics-like phenomena and complex pairing patterns~\cite{modawi1997some,paananen2006pairing,schafer2007atomic,o2011realizing,adams2012strongly,kurkcuoglu2018color,tajima2019quantum}. Pioneering experiments showed that unpolarized (\emph{i.e.} spin-population balanced) gases of three-component fermions exhibit interesting decay behavior related to the Efimov effect~\cite{ottenstein2008collisional,huckans2009three,williams2009evidence,wenz2009universal,lompe2010radio}.
However, density-dependent losses in those spatially inhomogeneous gases were unavoidably coupled to particle transport and heating, making the interpretation of those experiments challenging. The metastability of the three-component Fermi gas has yet to be comprehensively understood. 

In this work, we create box-trapped, uniform gases of three-component fermions with controllable spin-population imbalance and study their stability (Fig.~\ref{FIG:1}A). Such spatially homogeneous gases generally obey simpler dynamics than their inhomogeneous counterparts (see e.g.~\cite{eigen2017universal,bause2021collisions,ji2022stability,navon2021quantum}). 
This has enabled us to observe a surprising violation of the generic expectation that loss rates among spin components should be equal in a three-body process involving all three components. We rule out two credible explanations for this effect, indicating that unexpected physics is at play. 

\begin{figure*}[tb]
\includegraphics[width=2\columnwidth]{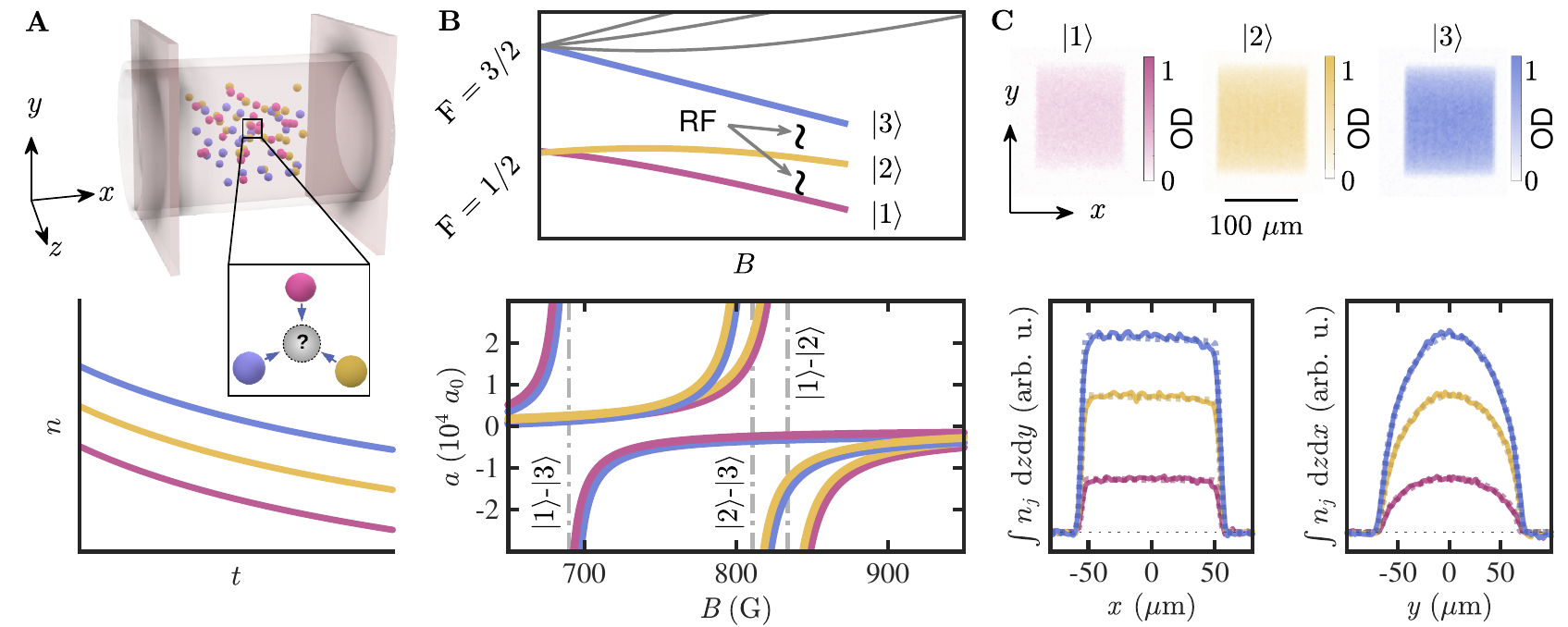}
\caption{Preparation of a homogeneous three-component Fermi gas. \textbf{(A)} \emph{Top:} Sketch of the optical box. \emph{Bottom:} The stability of the mixture is studied by measuring the density of each spin population over time.
 \textbf{(B)} \emph{Top:} Breit-Rabi diagram of the lowest hyperfine states of $^6$Li. The three (pseudo)-spins are encoded in the three lowest states, respectively $\ket{1}$, $\ket{2}$ and $\ket{3}$. The polarization of the three component mixture is controlled via radio-frequency (RF) pulses. \emph{Bottom:} Scattering length $a$ for each pair of spin states in units of $10^4 a_0$, where $a_0$ is the Bohr radius (adjacent colors match the corresponding pair of states). Vertical dashed lines show the locations of Feshbach resonances~\cite{zurn2013precise}. \textbf{(C)} \emph{Top:}  \emph{In-situ} absorption images of a typical polarized three-component mixture (averaged over 10 realizations), taken along the $z$ axis; the color scale corresponds to the optical density (OD). The length and radii of this conical box are $L = 120(2)~\mu$m, $R_1 = 75(1)~\mu$m and $R_2 = 73(1)~\mu$m, and its trap depth is $U_\text{box} = k_\text{B}\times$ 1.6(2)~$\mu$K, where $k_\text{B}$ is Boltzmann's constant. Note that boxes of various sizes were used in this work. \emph{Bottom:} Integrated density along the $x$ and $y$ axes for each component with fits to homogeneous density profiles (dotted lines)~\cite{SuppMat}. }
\label{FIG:1}
\end{figure*}

Our experiment begins with a gas of $^6$Li atoms in a red-detuned crossed optical dipole trap. The gas is prepared in an incoherent mixture of two of the three lowest Zeeman sublevels, which encode the three (pseudo-)spin components of our `spin-1' fermions (respectively labelled $\ket{1}$, $\ket{2}$, and $\ket{3}$,
see top panel of Fig.~\ref{FIG:1}B). This two-component mixture is evaporatively cooled at a bias magnetic field $B_0$ (which depends on the two states used) and then loaded into a blue-detuned optical box trap of wavelength $639$~nm.

We ramp the magnetic field to the field of interest $B$ in $200~\text{ms}$, which is slow compared to the two-body collision rate but fast compared to the lifetime of the two-component mixtures in the range of fields investigated~\cite{SuppMat}. We let the magnetic field settle for $100$~ms and then prepare the spin mixtures by sequentially applying variable radio-frequency (RF) pulses  to drive the $\ket{1}-\ket{2}$ and $\ket{2}-\ket{3}$ transitions (top panel of Fig.~\ref{FIG:1}B). For the range of $B$ explored in this work (deep in the Paschen-Back regime for $^6$Li), all three spins are nearly identically levitated against gravity using a magnetic field gradient~\cite{footnoteMagneticLevitation}. The fermions interact via binary contact interactions, characterized by an s-wave scattering length $a$ for each pair of spin states at a given magnetic field; each pair has a broad Feshbach resonance (see bottom panel of Fig.~\ref{FIG:1}B).

We hold the gas for a variable time $t$ before measuring the number of atoms of spin $\ket{j}$, $N_j$.
Typical \emph{in-situ} absorption images of each spin population in an imbalanced mixture are shown in Fig.~\ref{FIG:1}C (top), together with integrated column densities along the two directions $x$ and $y$ (bottom). 
The integrated density profiles of the three components are consistent with uniform densities.

\begin{figure}[!bt]
\includegraphics[width=\columnwidth]{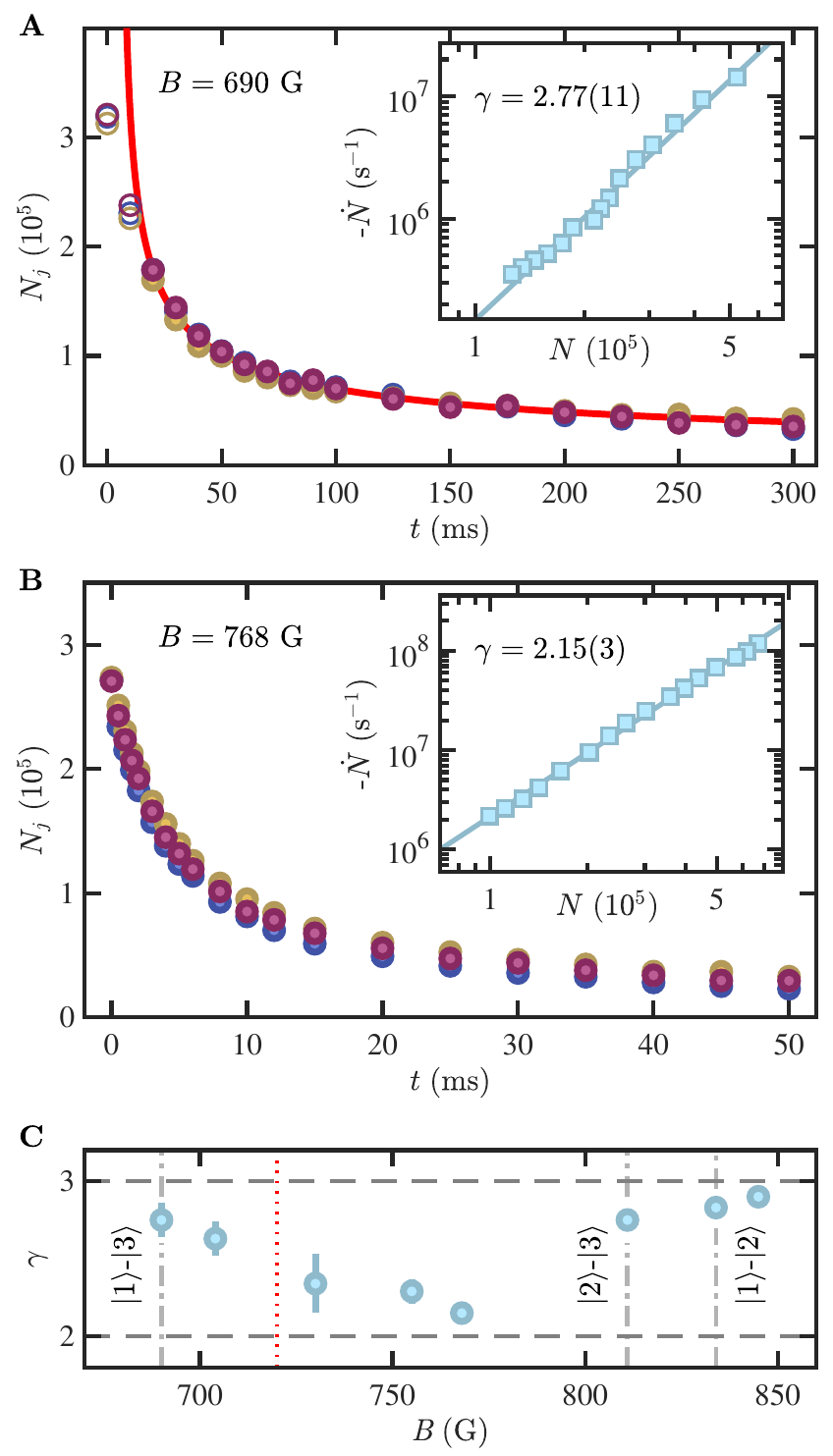}
\caption{Stability of the unpolarized three-component Fermi gas. \textbf{(A-B)} Typical decay at the $\ket{1}$-$\ket{3}$ resonance (\textbf{A}) and at $B=768~$G (\textbf{B}). The red line is a fit to Eq.~(\ref{EQ:1}). Early times (open symbols) are excluded from our analysis, see~\cite{SuppMat}. \emph{Insets:} Total loss rate $-\dot{N}$ versus $N$. The solid blue lines are power-law fits yielding $\gamma$.
\textbf{(C)} Exponent $\gamma$ versus magnetic field $B$. The Feshbach resonances are shown as dot-dashed gray lines. For $B$ larger than the red dotted line, the $\ket{12}$ and $\ket{23}$ Feshbach dimers can remain trapped (see text). Error bars on $\gamma$ displayed are only statistical (estimated by bootstrapping). For most points, the dominant source of uncertainty is an additional $10\%$ systematic error coming from box volume calibration~\cite{SuppMat}. The uncertainty in $B$ is smaller than the point size. }
\label{FIG:2}
\end{figure}

We first study the stability of spin-balanced samples, $N_1\approx N_2 \approx N_3$ as a function of magnetic field $B$. We typically evaporate a balanced mixture of $\ket{1}$-$\ket{3}$ near its Feshbach resonance at $B_0\approx 690$~G, ending with typically $N_1\approx N_3\approx 5\times10^5$ at $T/T_{\text{F}} \approx 0.25$ (where $T_\text{F}\approx400$~nK is the Fermi temperature) prior to ramping the field and creating the three-component mixture. 
In Fig.~\ref{FIG:2}A-B, we show two typical decays,  
from which we make two key observations: (i) the mixtures remain unpolarized during decay~\cite{balancedexpt}, and (ii) these losses involve all three states, as all two-component mixtures are much more stable~\cite{SuppMat}.  

The density uniformity of our samples enables us to model-independently characterize the three-component gas decay dynamics. 
In the insets of Fig.~\ref{FIG:2}A-B, we plot the total atom loss rate $\dot{N}\equiv \text{d}N/\text{d}t$ as a function of the total atom number $N\equiv N_1+N_2+N_3$. 
For the magnetic fields explored, the data is well captured by a power law $\dot{N} \propto - N^\gamma$, where the fitted $\gamma$ is displayed in Fig.~\ref{FIG:2}C.  
For a homogeneous unpolarized gas with a total density $n = N/V$ and a constant volume $V$, this implies that $\dot{n} \propto -n^\gamma$. Importantly, the exponent $\gamma$ encodes information on the number of independent particles involved in the loss events~\cite{mehta2009general}.

For both $B\lesssim 720$~G and $B\gtrsim 810$~G, we observe $\gamma\approx 3$.  This is consistent with losses dominated by energy-independent recombination involving three distinguishable fermions (for which we expect $\dot{n}\propto - n^3$).

The intermediate range $B\approx 720 - 810$~G is more complicated, and $\gamma$ deviates from 3. In that region, the scattering lengths $a_{12}$ and $a_{23}$, respectively between states $\ket{1}-\ket{2}$ and $\ket{2}-\ket{3}$, are large and positive, so that the products of three-body recombination can remain trapped~\cite{petrov2003three,ji2022stability}. This allows for subsequent inelastic atom-dimer collisions (involving three distinguishable atoms), an effectively two-body loss process~\cite{nakajima2010nonuniversal,lompe2010atom,braaten2010efimov}. Our intermediate $2\lesssim \gamma\lesssim 3$ is qualitatively consistent with this competition of two- and three-body effects.  

Focusing on the regions dominated by three-body recombination, we now explore polarized three-component mixtures, \emph{i.e.} with spin-population imbalance.  

In Fig.~\ref{FIG:3}A, we show a typical decay of a polarized sample at a field $B=845~$G, above all three broad Feshbach resonances (Fig.~\ref{FIG:1}B). 
The losses $\Delta n_j\equiv n_j(t)-n_j^\text{initial}$ are equal for all spins (bottom panel of Fig.~\ref{FIG:3}A). This is unsurprising since we expect the loss rate equation to take the form 
\begin{equation}
    \dot{n}_j=-L_3 n_1 n_2 n_3,
    \label{EQ:1}
\end{equation}
where $L_3$ is the (density-independent) three-body recombination loss coefficient~\cite{L3notationNote}.  The data at 845~G agrees very well with the numerical solution of Eq.~(\ref{EQ:1}), shown as solid lines in Fig.~\ref{FIG:3}A; the initial densities $n_j$ are the only fit parameters, $L_3$ being fixed to the value we measure from the decay of the unpolarized gas (see the caption of Fig.~\ref{FIG:3}).

\begin{figure*}[bt]
\includegraphics[width=2\columnwidth]{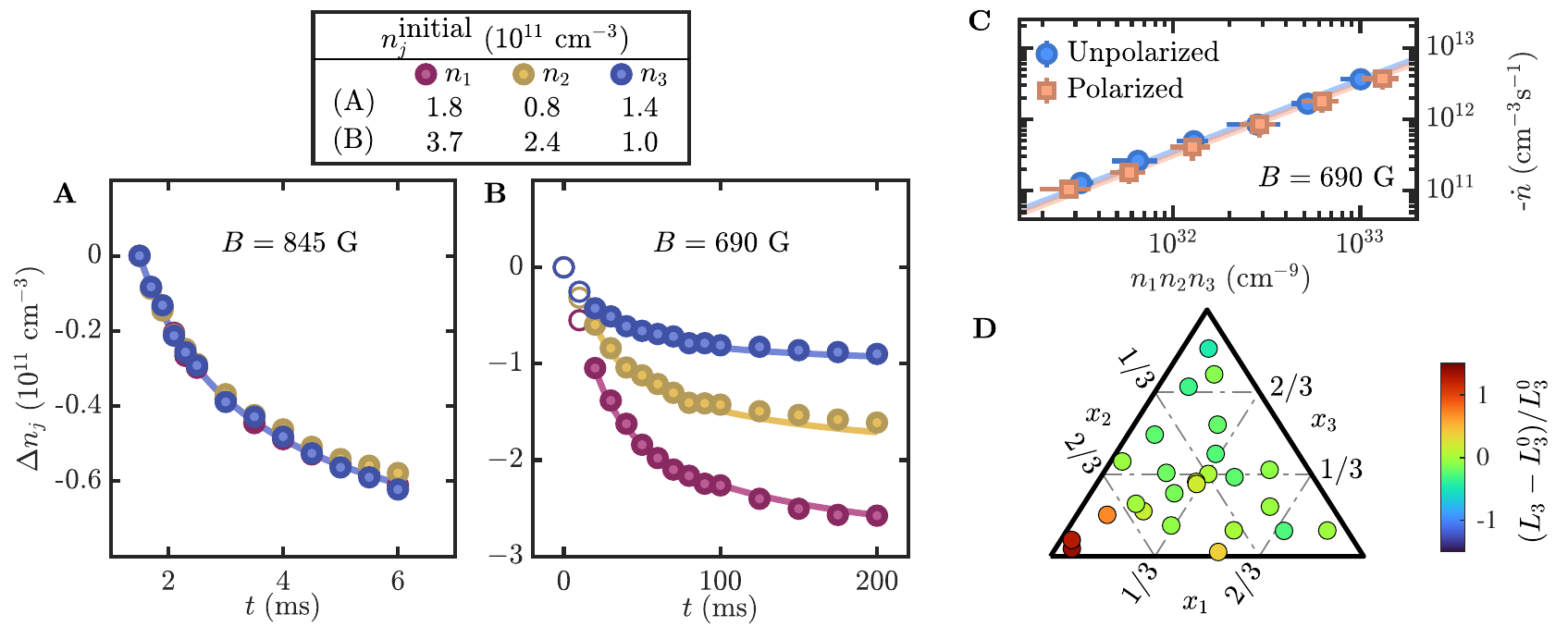}
\caption{Stability of the polarized three-component Fermi gas. \textbf{(A)} Decay of a polarized gas at $845$~G. Solid curves are fits to Eq.~(\ref{EQ:1}), with $L_3$ fixed to the value measured in the unpolarized decay: $L_3 (B = 845~\text{G}) = 2.4(1) \times 10 ^ {-20}~\text{cm}^6/\text{s}$ ($t=0$ corresponds to the end of the RF pulses.) \textbf{(B)} Decay of a polarized gas at the $\ket{1}$-$\ket{3}$ resonance. Solid curve are fits to Eq.~(\ref{eq:dotn}), with early time points excluded (open symbols)~\cite{SuppMat}. $L_3$ is fixed to the value measured in the unpolarized decay: $L^0_3 = 1.1(1)\times10^{-21}~\text{cm}^6/\text{s}$. We estimate additional systematic uncertainties on measurements of $L_3$ of $20\%$ due to box volume calibration~\cite{SuppMat}.  \textbf{(C)} Total loss rate $\dot{n}$ versus $n_1 n_2 n_3$, for both unpolarized (blue) and polarized (orange) gases at $690$~G, averaged over many experiments. Colored bands are fits (with uncertainties) to the three-body loss equation for $\dot{n}$, from which $L_3$ is extracted. \textbf{(D)} Three-body loss coefficient $L_3$ at $690$~G as a function of polarization, relative to $L_3^0$, plotted on a barycentric ternary plot~\cite{L30comment}.}
\label{FIG:3}
\end{figure*}

Surprisingly, the same measurement at the  Feshbach resonance of the $\ket{1}-\ket{3}$ states yields a qualitatively different outcome: the loss rates per spin are distinct, with the majority component showing larger absolute losses (Fig.~\ref{FIG:3}B).

Nonetheless, we observe that the \emph{total} loss rate still obeys the expected three-body rate equation $\dot{n} = -3L_3 n_1 n_2 n_3$ (orange points in Fig.~\ref{FIG:3}C).  What's more, the extracted $L_3$ matches the unpolarized gas value (blue points).

We systematically extract $L_3$ at 690~G as a function of polarization, see  Fig.~\ref{FIG:3}D.
Unlike its spin-1/2 counterpart~\cite{zwierlein2006fermionic,partridge2006pairing,nascimbene2009collective}, the polarization of the three-component gas is characterized by two parameters. The \emph{spin fractions} $x_j\equiv n_j/n$ provide a convenient parameterization, so that the polarization state can be represented on a ternary plot. 
We find that $L_3$ agrees with the unpolarized-gas value for nearly all polarizations, provided the spin fraction of $\ket{2}$ is not very high~\cite{SuppMat}.

\begin{figure*}[bt]
\includegraphics[width=2\columnwidth]{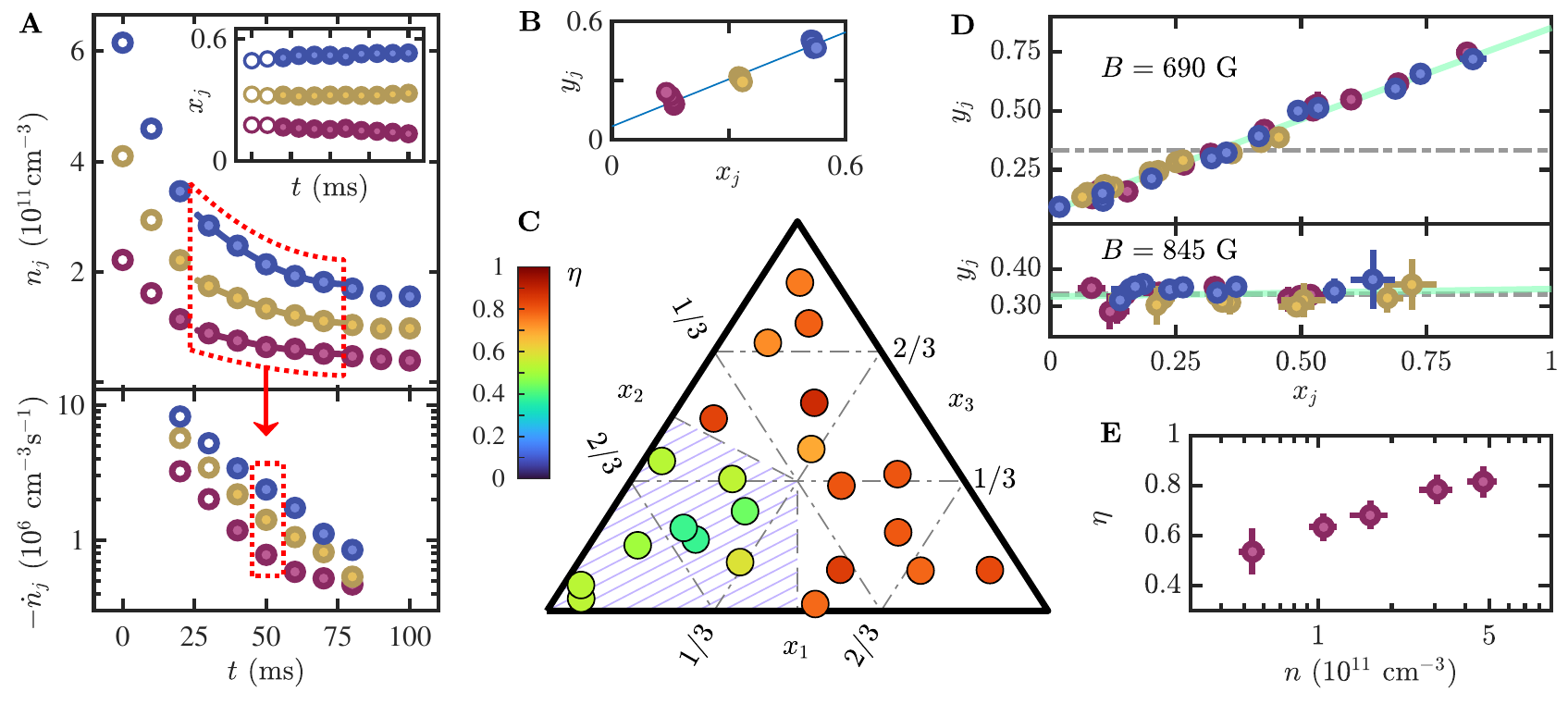}
\caption{Asymmetric losses in a polarized three-component Fermi gas. \textbf{(A)} \emph{Top:} Sketch of the extraction of spin fractions and relative losses $(x_j,y_j)$ from the $n_j(t)$ series. \emph{Inset:} $x_j$ versus time (see also~\cite{SuppMat}). \emph{Bottom}: $\dot{n}_j$ versus time. \textbf{(B)} The pairs $(x_j,y_j)$ extracted from the bottom of panel \textbf{A} lie on a line, whose slope defines $\eta$. \textbf{(C)}  $\eta$ at $690$~G as a function of polarization on a ternary plot. Each point is obtained by averaging over a single decay, as shown in \textbf{B}. The hatching marks two distinct regions in polarization space (see also~\cite{SuppMat}).
\textbf{(D)} \emph{Top:} $(x,y$) from the large (unhatched) region of the ternary plot. The green solid line is a linear fit, $\eta=0.80(3)$. The gray dash-dotted line shows $\eta = 0$. \emph{Bottom}: The same procedure at 845~G yields $\eta = 0.02(3)$.  We estimate an additional $5\%$ systematic error on $\eta$~\cite{SuppMat}. \textbf{(E)} $\eta$ versus total density $n$ at 690~G. 
The points are obtained by gathering data from several decays into density bins, then extracting $\eta$ (as in \textbf{D}). The highest two bins correspond to the data of \textbf{D}; this plot includes additional data at lower density. The variation of $\eta(n)$ is slow enough that it can be neglected over the density range considered during a typical decay (see \textbf{A}).}
\label{FIG:4}
\end{figure*}

With the total loss scaling established, we now investigate how losses are apportioned among the spins. As shown in Fig.~\ref{FIG:4}A, we see that $\dot{n}_j$ appear proportional to their respective spin fractions $x_j$. To characterize this relation, we introduce the \emph{relative loss} $y_j\equiv\dot{n}_j/\dot{n}$ and compare it directly to the $x_j$, as shown in Fig.~\ref{FIG:4}B. 

This measurement suggests a simple relation: 
\begin{equation}
    y_j=\eta x_j+\frac{1-\eta}{3},
\end{equation}
where the proportionality of the losses is captured by the \emph{asymmetry} $\eta$. The constant $(1-\eta)/3$ is constrained by the definition of the parameters $x_j$ and $y_j$; $\eta = 0$ corresponds to symmetric losses as in Eq.~(\ref{EQ:1}).

We now explore $\eta$ at 690~G as a function of polarization. Following the procedure outlined in Fig.~\ref{FIG:4}A-B, the ternary plot is populated with the values of $\eta$ extracted from each time series. 

We observe that the polarization space is separated into two regions. In the (larger) region where $\ket{1}$ or $\ket{3}$ is the largest component (unhatched region in Fig.~\ref{FIG:4}C), $\eta$ is robust to spin composition, regardless of whether $\ket{2}$ is the median or the minority component. In the (smaller) region where $\ket{2}$ is the largest component (hatched in Fig.~\ref{FIG:4}C), $\eta$ is smaller, although distinctly non-zero. Note that exchanging spins $\ket{1}$ and $\ket{3}$ leaves Fig.~\ref{FIG:4}C essentially unchanged, which reflects the approximate symmetry of the three scattering lengths (see Fig.~\ref{FIG:1}B).

Gathering the data from the larger (unhatched) region, we observe that the relative loss law appears indeed to be universal over a large range of spin fractions. Fitting the data, we find $\eta = 0.80(3)$ (see caption of Fig.~\ref{FIG:4}D).

On the other hand, applying the same procedure at $B=845$~G we find $\eta = 0.02(3)$ (Fig.~\ref{FIG:4}D), characteristic of the usual three-body loss dynamics of Eq.~(\ref{EQ:1}).

The behavior of both the total and relative losses suggests that the three-component mixture at $690$~G obeys the following decay dynamics:
\begin{align} \label{eq:dotn}
    \dot{n}_j = -3 L_3 \left(\eta x_j + \frac{1-\eta}{3}\right)\ n_1 n_2 n_3.
\end{align}
In Fig.~\ref{FIG:3}B, we show as solid lines the solution of Eq.~(\ref{eq:dotn}) with fixed $L_3$ and $\eta$, and only the initial $n_j$ as adjustable parameters; the agreement with the experimental data is excellent.

Gathering decay measurements across a large density range, we observe that $\eta$ is weakly density dependent (Fig.~\ref{FIG:4}E), although this variation is slow enough so that it is negligible over our typical decay series.

We now turn to hypotheses for explaining this anomalous loss asymmetry. First, note that the rate equation Eq.~(\ref{EQ:1}) is an approximation; more fundamentally, losses are expressed as correlators of quantum fields~\cite{kagan1985effect}. 
Calculating $\dot{n}_j$ for a generic model of a gas on a lattice in the Lindblad formalism yields~\cite{SuppMat,FelixYvan}
\begin{equation}
    \dot{n}_j \propto  -\sum_\alpha \left \langle a_{3,\alpha}^\dagger a_{2,\alpha}^\dagger
    a_{1,\alpha}^\dagger
    a_{1,\alpha} a_{2,\alpha} a_{3,\alpha} \right \rangle
    \label{EQ:Lindblad}
\end{equation}
where $a_{j,\alpha}$ (resp. $a^\dag_{j,\alpha}$) is the annihilation (resp. creation) operator for state $\ket{j}$ at lattice site $\alpha$, under the assumptions that no spin-changing collisions occur and that losses are due to local three-body recombination.  As the right-hand side does not depend on the spin state, $\eta = 0$ within this model. This conclusion holds even in the presence of nontrivial correlations that would preclude factoring Eq.~(\ref{EQ:Lindblad}) into a product of densities. 

Alternatively, this effect could originate from many-particle scattering processes. The form of Eq.~(\ref{eq:dotn}) is indeed qualitatively reminiscent of so-called \emph{avalanche} mechanisms, in which secondary collisions occur between the products of three-body recombination and other atoms in the gas, resulting in excess losses that augment the loss rate prefactor with spin-fraction-dependent terms. Avalanche losses have been a debated topic in ultracold few-body dynamics~\cite{schuster2001avalanches,zaccanti2009observation,langmack2013avalanche,machtey2012universal,hu2014avalanche,zenesini2014resonant}.

Under an avalanche process, one would expect that the number of excess particles of spin $\ket{j}$ lost should be proportional to the fraction of atoms in that spin state, \emph{i.e.} equal to $c x_j$. The prefactor $c$ should be independent of the spin state to ensure that an unpolarized gas remains unpolarized during decay; $c$ is then the average number of excess particles per recombination event. In that case, $\dot{n}\propto-(3+c)n_1 n_2 n_3$.

In this model, $\eta=c/(3+c)$~\cite{SuppMat}, and our largest measured $\eta$ would imply $c \gtrsim 12$. However, a generic kinetic model~\cite{zaccanti2009observation} indicates that $c \approx 4$ given the binding energies of the $\ket{12}$ and $\ket{23}$ Feshbach dimers. Furthermore, repeating this experiment in a box whose size is about the mean free path or smaller, we find the same asymmetry~\cite{SuppMat}. This further rules out an avalanche scenario. Finally, evaporation in a collisionally dense gas could give rise to asymmetric losses, as suggested in~\cite{parish2009evaporative}, but is also essentially ruled out as varying the box depth does not affect the loss rate~\cite{SuppMat}.

Future work should elucidate this puzzle, for example by measuring the energy dynamics and temperature dependence of this process, as well as by characterizing the asymmetry across interaction regimes.
This work also opens several new avenues, such as studying prethermalization dynamics of the metastable mixture~\cite{ueda2020quantum,huang2020suppression} and searching for signatures of three-component pairing correlations at low temperature~\cite{modawi1997some,paananen2006pairing}.

We thank Chris Greene, Jose d'Incao, Leonid Glazman, Steve Girvin, Carlos S{\'a} de Melo, Alexander Schuckert, and Francesca Ferlaino for helpful discussion. We thank Yvan Castin, Félix Werner, Frédéric Chevy, and Zoran Hadzibabic for critical comments on the manuscript. This work was supported by the NSF (Grant Nos. PHY-1945324 and PHY-2110303), DARPA (Grant No. W911NF2010090), the David and Lucile Packard Foundation, and the Alfred P. Sloan Foundation. J.T.M. acknowledges support from the Yale Quantum Institute. G.L.S acknowledges support from the NSF Graduate Research Fellowship Program.

\newpage
\cleardoublepage
\setcounter{figure}{0}
\setcounter{equation}{0}
\renewcommand{\thefigure}{S\arabic{figure}}
\renewcommand{\theequation}{S\arabic{equation}}

\makeatletter
\def\maketitle{
\@author@finish
\title@column\titleblock@produce
\suppressfloats[t]}
\makeatother

\makeatletter
\def\maketitle{
\@author@finish
\title@column\titleblock@produce
\suppressfloats[t]}
\makeatother

\onecolumngrid
\begin{center}
\large{\bf{Supplementary Material\\Observation of Anomalous Decay of a Polarized Three-Component Fermi Gas}}
\end{center}

\section{I. Characterizing Optical Boxes}

\begin{figure}[H]
\centering
\includegraphics[width=0.9\columnwidth]{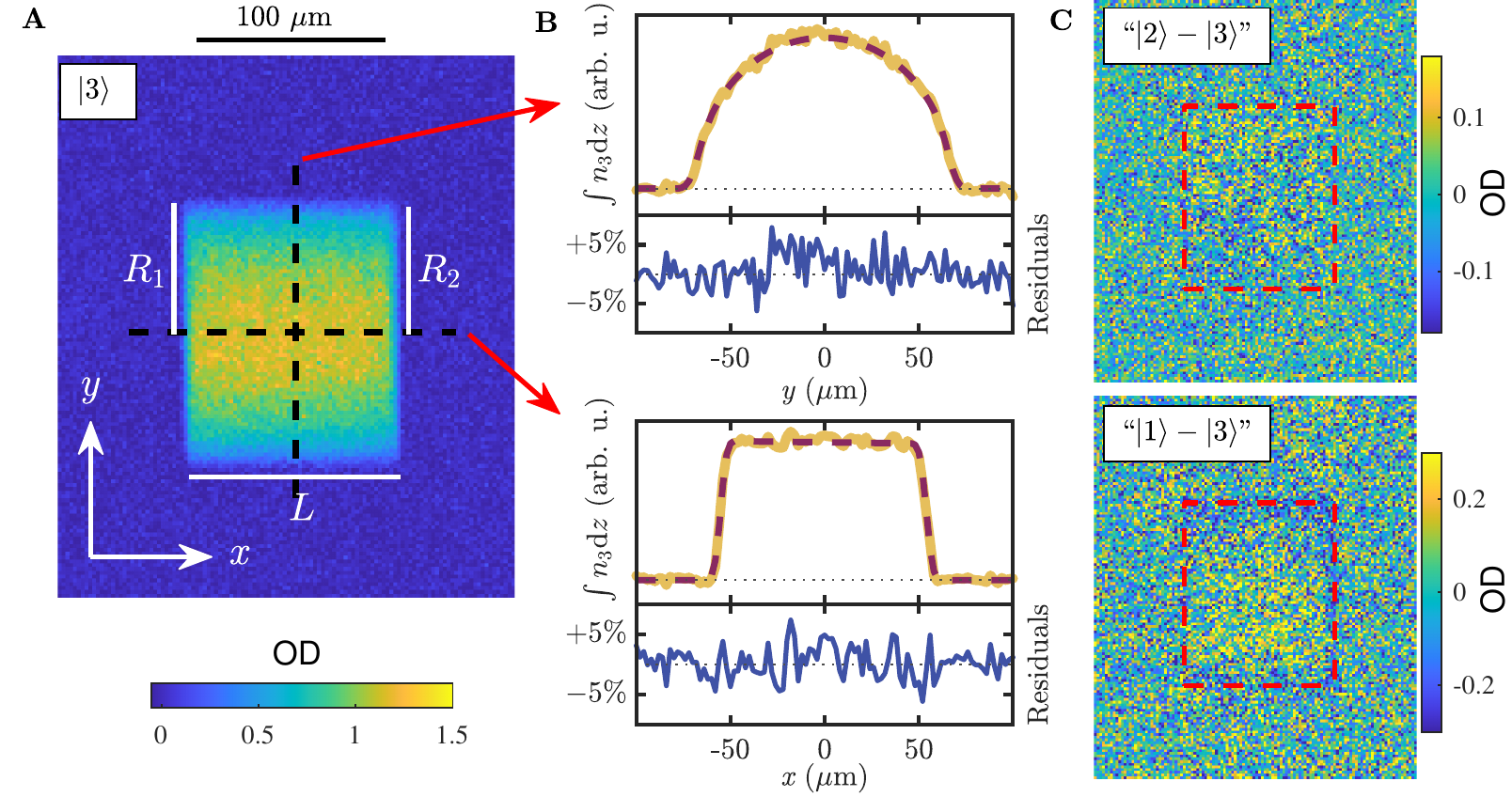}
\caption{Homogeneity of a polarized three-component mixture in the optical box. \textbf{(A)} Example of an \emph{in-situ} optical density image of spin $\ket{3}$ (same data as Fig.~1C, with $x_1 = 0.16$, $x_2 = 0.35$, and $x_3 = 0.49$). \textbf{(B)} \emph{Top}: Cuts of the density profile (yellow, solid) and the fit (purple, dashed) along the $y$ (top) and $x$ (bottom) axes (dashed black lines in \textbf{A}). The fit to the OD image is a smoothed cone (see text). Residuals (normalized to the peak fitted density along each cut) are shown underneath.  
\textbf{(C)} \emph{Top}: Subtracting the measured \emph{in-situ} density of $\ket{3}$ from $\ket{2}$ (after scaling $\ket{2}$ to have the same mean density as $\ket{3}$). OD scale is the same as \textbf{A} and the dashed red rectangle indicates the region of the box. \emph{Bottom}: Likewise subtracting spin $\ket{3}$ from $\ket{1}$.  }
\label{SFIG:trapinsitu}
\end{figure}

We determine the volume of the optical box of Fig.~1 of the main text by fitting the \emph{in-situ} density profiles $\int\text{d}z\;n_j(x,y,z)$ (where $z$ is the imaging line of sight, Fig.~\ref{SFIG:trapinsitu}A) with a conical model of radii $R_1$ and $R_2$, and length $L$, convolved with a Gaussian function of standard deviations $\sigma_x$ and $\sigma_y$ (that capture imperfections from finite resolution of the imaging and box projection systems). We find $R_1= 75(1)\ \mu$m, $R_2= 73(1)\ \mu$m, $L=120(2)\ \mu$m. The values of $\sigma_x= 2.4\ \mu$m and $\sigma_y=3.8\ \mu$m reflect the slightly different quality of projection of the `end caps' and the `tube' of the box. In Fig.~\ref{SFIG:trapinsitu}B, we display density profile cuts of the box of Fig.~1C projected along the $x$ and $y$ axes together with the fits (dashed purple lines). 

If imperfections were entirely due to the box projection system - a conservative assumption - we estimate that $\approx90\%$ of the atoms are within $10\%$ of the average density in the box. Additionally, we observe during a typical decay that the volume decreases by $\approx 10\%$ when the atom number decreases by a factor of $\approx 5$; taking this small effect into account as an uncertainty on the volume, we estimate an uncertainty of 10\% on $\gamma$ and of 20\% on $L_3$.  This also contributes to uncertainties on $x_j,y_j$ of $\approx 1\%$ and $\approx 5\%$ respectively (reflecting the difference between fitting $N$ and $\dot{N}$ rather than $n$ and $\dot{n}$).  Ultimately we find that the uncertainty in $\eta$ due to this effect is $\approx 5\%$.

In Fig.~\ref{SFIG:trapinsitu}C, we compare the density profiles of the three spin states by scaling and subtracting the profile of $\ket{3}$ from the others.  We see that the density profiles are essentially identical up to rescaling.

\vspace*{30px}

\section{II. Two- versus three-component mixtures decay}

\begin{figure}[H]
\centering
\includegraphics[width=0.9\columnwidth]{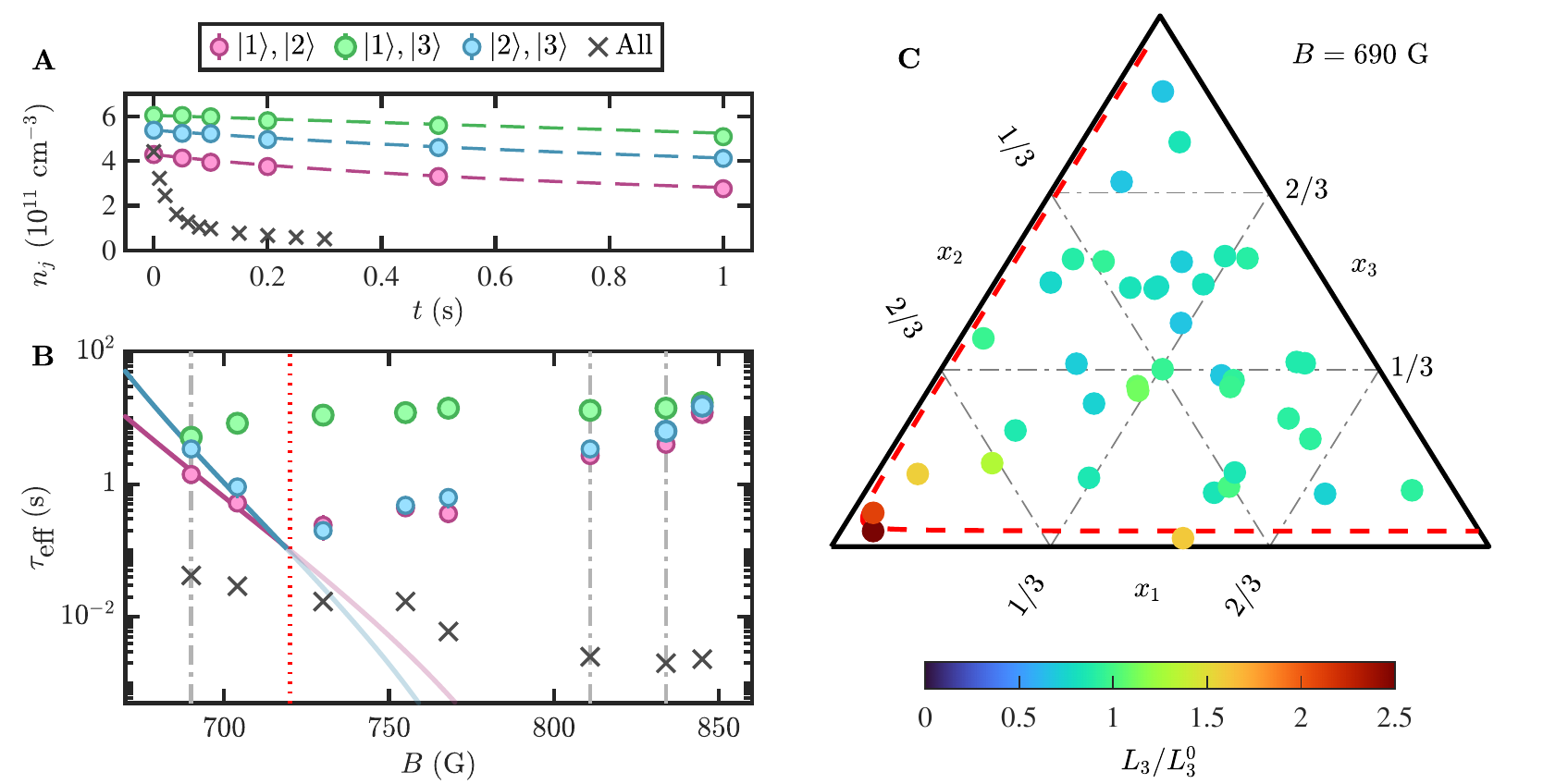}
\caption{Stability of two- versus three-component mixtures. \textbf{(A)}  Decay of unpolarized two-component gases at 690~G; a three-component gas decay is shown as black crosses for reference. \textbf{(B)} Effective timescales across magnetic fields. Feshbach resonances are indicated by vertical dash-dotted lines.  At each field shown, the two-component gases have densities (per spin state) equal to or greater than those of the corresponding three-component gas.  Solid pink and blue curves show predictions from universal recombination to Feshbach dimers~\cite{ji2022stability,petrov2003three}. For $B$ larger than the red dotted line, the $\ket{12}$ and $\ket{23}$ Feshbach dimers can remain trapped. \textbf{(C)} $L_3$ versus polarization at 690~G (normalized to $L_3^0$ as in the main text), includes additional data at lower total density ($n\sim 5\times10^{10} ~\text{cm}^{-3}$) not shown in main text. Dashed red curve gives polarization outside which two-component processes are expected to account for 10\% of total losses or more.}
\label{SFIG:twocomplimit}
\end{figure}

We show in this section that in the range of $B$ explored in this work, unpolarized three-component mixtures decay much faster than any two-component ones.
In Fig.~\ref{SFIG:twocomplimit}A, we show examples of decays of two-component mixtures (which may include evaporative losses in addition to intrinsic recombination processes). We compare two- and three-component decays quantitatively in Fig.~\ref{SFIG:twocomplimit}B by extracting an (early-time) effective timescale for losses $\tau_\text{eff} = n/|\dot{n}|$. The black crosses correspond to three-component mixtures, with a density per spin state in the range $n_j\approx 1.5 - 4\times 10^{11}~$cm$^{-3}$. In green, blue and pink, we show $\tau_\text{eff}$ for the two-component mixtures; for each field, the  density per spin state is up to a factor of 4 larger than the corresponding three-component data, but never smaller. Even in that case, $\tau_\text{eff}$ is at least an order of magnitude larger than the corresponding three-component one.

However, this will no longer hold for extremely polarized three-component gases.
At $B=690~$G, both $a_{12}$ and $a_{23}$ are positive so that the corresponding two-component mixtures are already unstable with respect to three-body recombination (involving two identical fermions). Furthermore, the corresponding binding energies of the $\ket{12}$ and $\ket{23}$ dimers are such that all recombination products escape from the box (see section VII). Thus, a mixture of spins $\ket{i}-\ket{j}$ would decay following a recombination law $\dot{n}_j = - L_3^{ij}(2n_j^2 n_i + n_i^2 n_j)/3$~\cite{petrov2003three,ji2022stability}. We note that the two-component recombination rates agree with the universal prediction of~\cite{petrov2003three} (solid lines of Fig.~\ref{SFIG:twocomplimit}B) at fields below 720~G (red dotted line of Fig.~\ref{SFIG:twocomplimit}B).  At higher magnetic fields, the formation of $\ket{12}$ and $\ket{23}$ dimers does not directly lead to losses, as their binding energies are less than $ 6U_\text{box}$~\cite{ji2022stability}.

To estimate the influence of two-component losses on the main text's analysis, we calculate the ratio of the total losses due to three-component recombination ($\dot{n}_\text{three}\equiv -3L_3 n_1 n_2 n_3$) to two-component ones ($\dot{n}_\text{two}\equiv -L_3^{12} (n_1^2 n_2 + n_2^2 n_1) - L_3^{23} ( n_2^2 n_3 + n_3^2 n_2)$):
\begin{equation}
    \frac{\dot{n}_\text{two}}{\dot{n}_\text{three}} = \frac{L_3^{12}}{3 L_3} \left(1 / x_3 - 1 \right) +  \frac{L_3^{23}}{3 L_3} \left(1 / x_1 - 1 \right)
    \label{EQ:twobodythreebodyratio}
\end{equation}
Extracting $L_3^{12}$ and $L_3^{23}$ from Fig.~\ref{SFIG:twocomplimit}A, we find ratios $L_3^{12}/(3 L_3) < 0.003$ and $L_3^{23}/(3 L_3) < 0.001$. Setting a limit of $\dot{n}_\text{two}/\dot{n}_\text{three}<  10\%$, we find the polarization limit depicted in Fig.~\ref{SFIG:twocomplimit}C. For mixtures less polarized than this limit, three-component losses are dominating, and we ignore losses arising from two-component processes.

\section{III. Early-time Decay}

In the main text, we excluded early times from the analysis of decays at $B=690~$G. Here we show in more detail that the early-time behavior at that field is distinct from behavior at later times.

\begin{figure}[H]
\centering
\includegraphics[width=0.9\columnwidth]{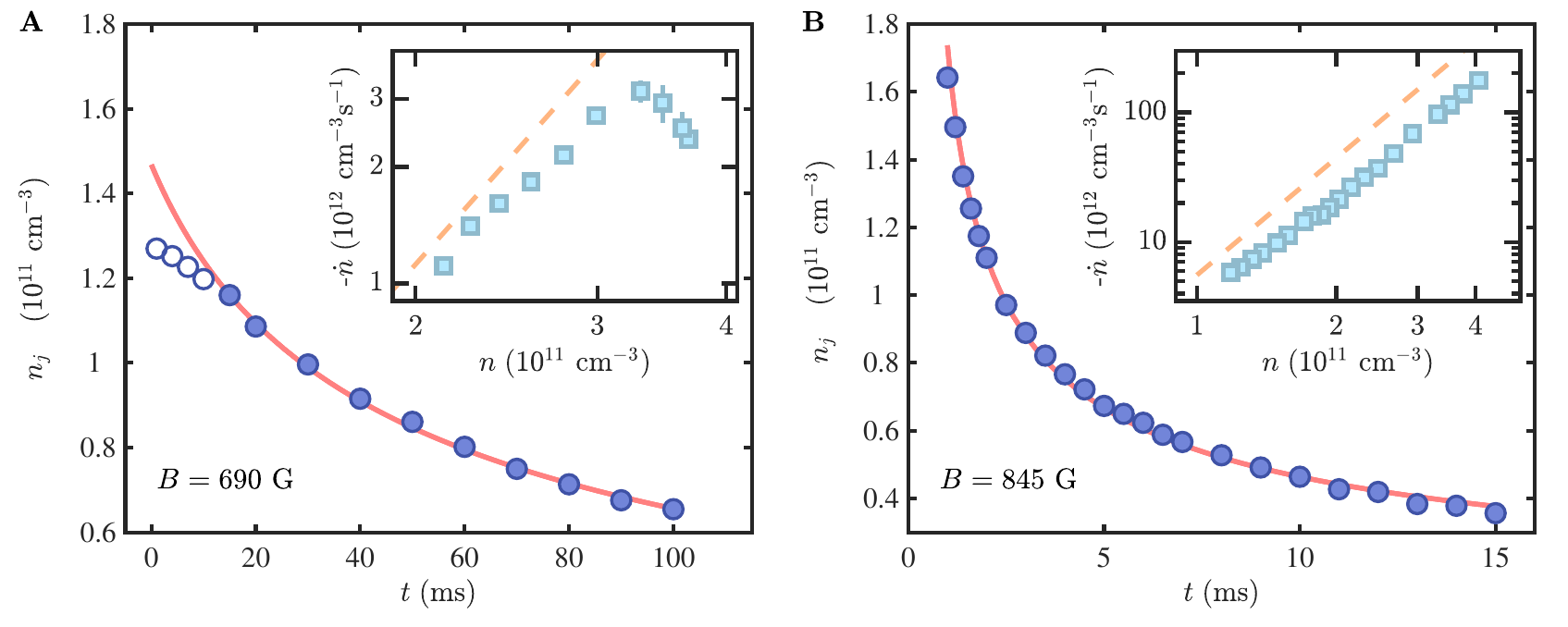}
\caption{Early-time decay of unpolarized three-component gases at $B=690~$G {\bf (A)} and $B=845~$G {\bf (B)}.  Blue points are averages of the three components, and the solid red lines are three-body fits for $t\geq15~$ms and for the entire time range in {\bf A} and {\bf B} respectively.  \emph{Insets}: $\dot{n}$ versus $n$. Dashed lines are guides to the eye showing the $n^3$ scaling on log-log scales.}
\label{SFIG:earlytime}
\end{figure}

In Fig.~\ref{SFIG:earlytime}A, we show a decay of an unpolarized sample at $690$~G. We see that for $t\lesssim 10~$ms, the decay is slower than at later times (see red solid line in Fig.~\ref{SFIG:earlytime}). This effect is even more obvious for $\dot{n}$ versus $n$ (inset of Fig.~\ref{SFIG:earlytime}A). Only at later times the data follows a $\gamma\approx 3$ law. This effect is also present in polarized-gas measurements, but for $t\gtrsim 15$~ms the data is well described by a single $\gamma\approx 3$ for all our polarizations. By contrast, for the faster losses at $B\gtrsim 845~$G (see Fig.~\ref{SFIG:earlytime}B), the data follows a simple $\gamma\approx 3$ law, even from the earliest measured times (inset of Fig.~\ref{SFIG:earlytime}B).

We speculate that this behavior is due to the protocol used to create the three-component mixtures, via global rotation of the spins. Indeed, for the slower losses at $B=690~$G, spin coherence at early times would affect both elastic and inelastic collision cross sections. It would be interesting to investigate the interplay between the coherent RF preparation, inelastic losses and possible Zeno effects in this system.

\section{IV. Spin Fraction Dynamics}

\begin{figure}[H]
\centering
\includegraphics[width=0.9\columnwidth]{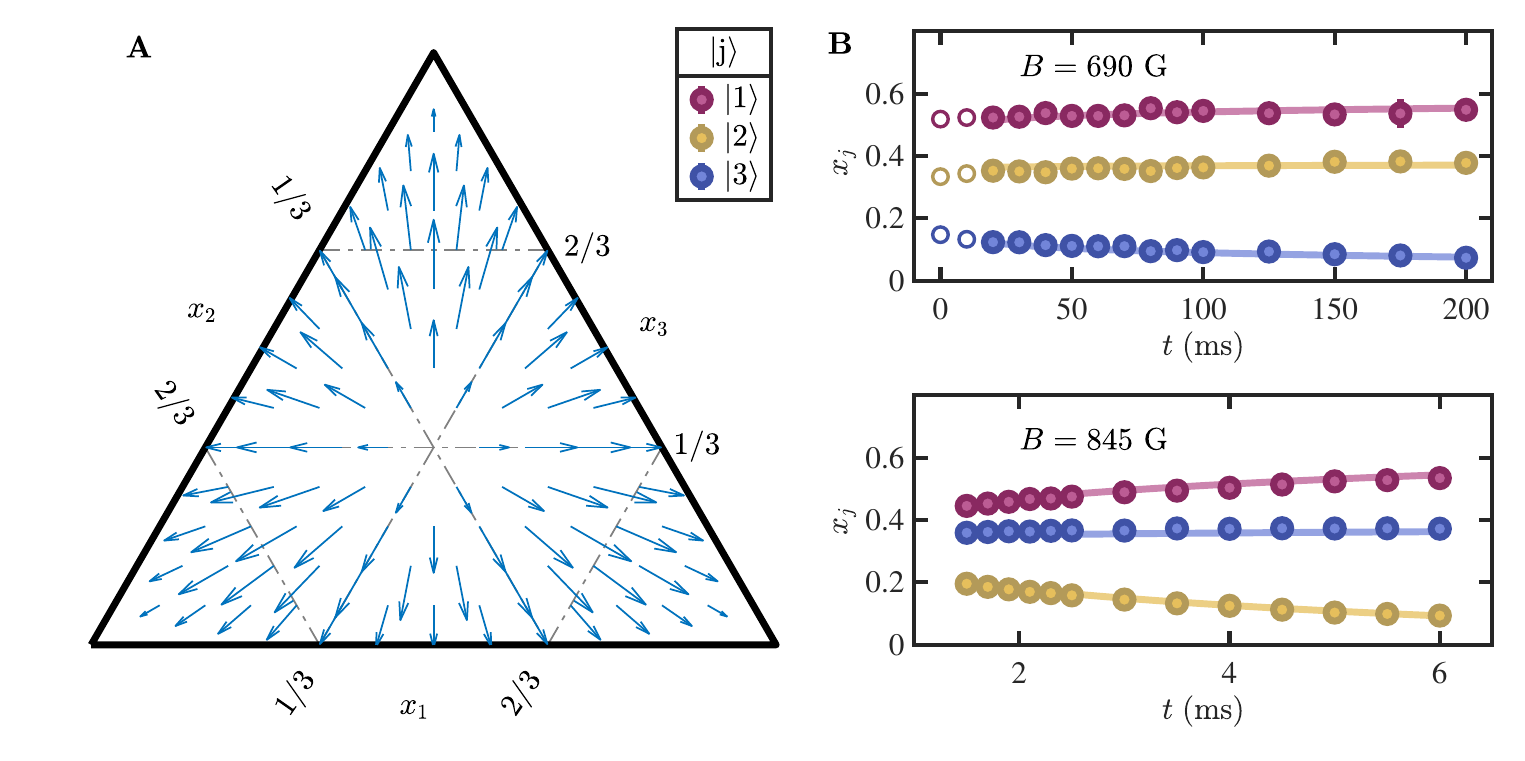}
\caption{Spin fraction dynamics under asymmetric three-body decay. \textbf{(A)} Map of the polarization flow.  \textbf{(B)} Examples of polarization dynamics at $B=690~$G (top) and $B=845~$G (bottom). Empty points are neglected due to early-time effects. 
The solid lines are solutions to Eq.~(\ref{EQ:polflow}).}
\label{SFIG:polarizationdynamics}
\end{figure}

Here we show that under the effective decay Eq.~(3) in the main text, the two distinct regions of polarization in the ternary plot Fig.~4C are closed under time evolution, \emph{i.e.} time dynamics does not allow for crossing from the hatched area to the unhatched one (and vice-versa). Indeed Eq.~(3) implies
\begin{align}
    \dot{\mathbf{x}} & = (1 - \eta) \frac{3 L_3 n_1 n_2 n_3}{n_1 + n_2 + n_3}
    \left (
    \mathbf{x} -
    \mathbf{c}
    \right),\label{EQ:polflow}
\end{align}
where $\mathbf{x} = (x_1, x_2, x_3)$ and $\mathbf{c}=\frac{1}{3}(1,1,1)$.
The spin fraction flow is a ray from the unpolarized state $\mathbf{c}$, shown in Fig.~\ref{SFIG:polarizationdynamics}A.  

In Fig.~\ref{SFIG:polarizationdynamics}B, we show the spin fractions over time, with the  predictions of Eq.~(\ref{EQ:polflow}) (solid lines). We note that the magnitude $|\dot{\mathbf{x}}|$, the `speed' of this flow, is suppressed for large $\eta$. Indeed, at $B=690~$G (top) the spin fractions barely change, while at $B=845~$G (bottom), they visibly vary.

This polarization map has an interesting  consequence. 
A two-component gas with a small contaminant of a third state will purify into a two-component mixture at a large cost in particle loss if $\eta$ is large.
Indeed, considering a small contaminant $x_2 \ll x_1 = x_3$, the loss rate equations simplify, yielding $(\dot{n}_1 + \dot{n}_3)/\dot{n}_2  \approx (2+\eta) / (1 - \eta )$. For instance, for $\eta=0.8$ we have $\dot{n}_1 + \dot{n}_3\approx 14 \dot{n}_2$, an inefficient way to get rid of the impurity state.

\section{V. RF Preparation Protocols}

\begin{figure}[H]
\centering
\includegraphics[width=0.9\columnwidth]{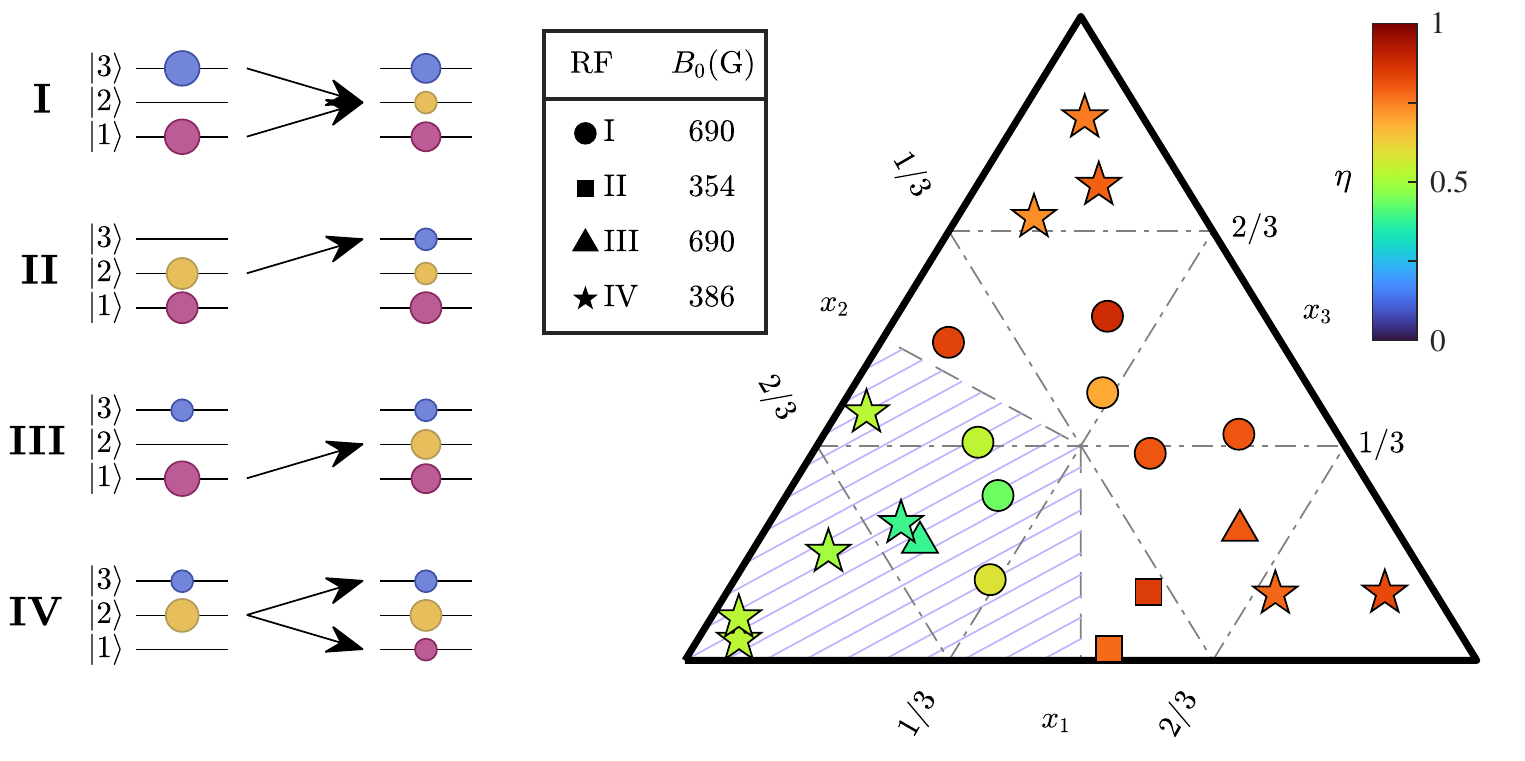}
\caption{Robustness of $\eta$ with respect to RF preparation protocols. Symbols in the ternary plot correspond to the protocols indicated in the legend.    
Colored circles indicate initial and final spin populations, and arrows denote the RF pulses. $B_0$ is the magnetic field at which the two-component gas was evaporated (prior to the RF pulses).}
\label{SFIG:protocols}
\end{figure}

Here, we provide further details on the RF preparation of the three-component mixtures. To cover extensively the polarization space, we used several RF protocols, depicted in Fig.~\ref{SFIG:protocols}. The ternary plot data in Fig.~\ref{SFIG:protocols} is populated with the data of Fig.~4C, with symbols corresponding to the protocol used. We see that the different protocols produce consistent values in the regions where they overlap. Depending on the spin states used, the initial evaporation of the two-component mixture was carried out at different magnetic fields $B_0$ as noted in Fig.~\ref{SFIG:protocols}.

\section{VI. A Simple Loss Model on a Lattice}

In this section we show that a generic model does not account for the loss asymmetry. We consider a model of a gas on a three-dimensional lattice, of lattice spacing $b$.

Treating the gas as an open quantum system, we calculate the rate of change of atom number $\frac{\text{d}} {\text{d}t} \braket{N_j} = \text{tr}\left( \dot{\rho} N_j \right)$ using the Lindblad equation~\cite{braaten2017lindblad,manzano2020short}:
\begin{equation}
    \dot{\rho} = \frac{1}{i\hbar} \left [ H_\text{sys}, \rho \right ] + \sum_\alpha \left( L_\alpha \rho L_\alpha ^\dagger - 1/2~ \{L_\alpha^\dagger L_\alpha, \rho \} \right ),
\end{equation}
where $\rho$ is the density matrix of the system, and $H_\text{sys}$ is the system's (conservative) Hamiltonian, containing the kinetic, potential and elastic interaction energies. The atom number operator in state $\ket{j}$ is defined as $N_j = \sum_\alpha b^3 a_{j,\alpha}^\dagger a_{j,\alpha}$~\cite{operatorDefNote}, where operators $a_{j,\alpha}$ and $a^\dag_{j,\alpha}$ respectively annihilate and create a particle of spin $\ket{j}$ at lattice site $\alpha$. These operators satisfy the anti-commutation relations: $\{a_{i,\alpha},a_{j,\beta}\}  = 0$ and
    $\{a_{i,\alpha},a_{j,\beta}^\dagger\} = \delta_{i,j}\delta_{\alpha,\beta}/b^3$.
$L_\alpha = \sqrt{\kappa} a_{1,\alpha} a_{2,\alpha} a_{3,\alpha}$ is the non-hermitian jump operator which represents the inelastic physics: three particles of different spin states scatter inelastically with a  probability~$\propto \kappa$ when they are located on the same lattice site $\alpha$, causing them to escape the trap. This yields
\begin{equation}
    \frac{\text{d}}{\text{d}t}\braket{N_j} = \sum_{\alpha,\beta} \braket{L_\alpha^\dagger (b^3 a_{j,\beta}^\dagger a_{j,\beta}) L_\alpha - 1/2 \{L_\alpha^\dagger L_\alpha, b^3 a_{j,\beta}^\dagger a_{j,\beta} \}}, \label{EQ:lindblad}
\end{equation}
 Since we have observed no spin changing collisions in any two-component mixture, we have assumed that $H_\text{sys}$ commutes with all $N_j$.
Using $\braket{L_\alpha^\dagger (b^3 a_{j,\beta}^\dagger a_{j,\beta}) L_\alpha}  = \braket{L_\alpha^\dagger L_\alpha (b^3 a_{j,\beta}^\dagger a_{j,\beta})} - \delta_{\alpha,\beta} \braket{L_\alpha^\dagger L_\alpha}$ and $\braket{\{L_\alpha^\dagger L_\alpha, (b^3 a_{j,\beta}^\dagger a_{j,\beta})\}}  = 2 \braket{ L_\alpha^\dagger L_\alpha (b^3 a_{j,\beta}^\dagger a_{j,\beta})}$, we simplify Eq.~(\ref{EQ:lindblad}) into
\begin{equation}
    \frac{\text{d}}{\text{d}t}\braket{N_j} = - \sum_\alpha \braket{L_\alpha^\dagger L_\alpha}
    = - \kappa \sum_\alpha \braket{ a_{3,\alpha}^\dagger a_{2,\alpha}^\dagger  a_{1,\alpha}^\dagger  a_{1,\alpha}  a_{2,\alpha}  a_{3,\alpha} }
\end{equation}

There is thus no dependence on the spin label $j$, and each spin component shares the losses equally: $\eta = 0$. 

\section{VII. Avalanche Losses}

In an avalanche process, subsequent elastic scattering following a recombination event leads to the loss of additional atoms. In this section, we elaborate on avalanches as a candidate for the origin of the asymmetric losses. We show that avalanches have two rather generic consequences, neither of which are observed in our experiment.

Under an avalanche process, one would expect that the number of excess particles of spin $\ket{j}$ lost should be proportional to the fraction of atoms in that spin state, \emph{i.e.} equal to $c x_j$. 
To ensure that an unpolarized gas remains unpolarized during decay, $c$ should be independent of the spin state; $c$ is then the average number of excess particles lost per recombination event. In that case, we expect $\dot{n}_j = -K_3 (1 + c x_j) n_1 n_2 n_3$, where $K_3$ is the event rate per unit volume. This equation qualitatively matches the asymmetric loss equation proposed in Eq.~(3), with $\eta = c/(3+c)$. For our measured value $\eta \approx 0.8$, this implies $c \approx 12$. 

On the other hand, we can estimate $c$ using a kinetic model~\cite{zaccanti2009observation}. We assume that three-body recombination produces a dimer of binding energy $\epsilon_\text{b}$ and a free atom. If all three incoming participants are at rest, the dimer will have a kinetic energy $\epsilon_\text{b}/3$ after formation.
In a subsequent elastic collision with a free atom at rest, the dimer will retain
$5/9$ of its pre-collision kinetic energy on average. If the gas is collisionally opaque to the dimer (though this is unlikely, see section VIII), these collisions will kick free atoms from the trap until the dimer has less energy than the trap depth. The dimer can thus kick out about $\log_{5/9}\left( 3 U_\text{box}/\epsilon_\text{b}\right)$ particles.

\begin{figure}[H]
\centering
\includegraphics[width=0.9\columnwidth]{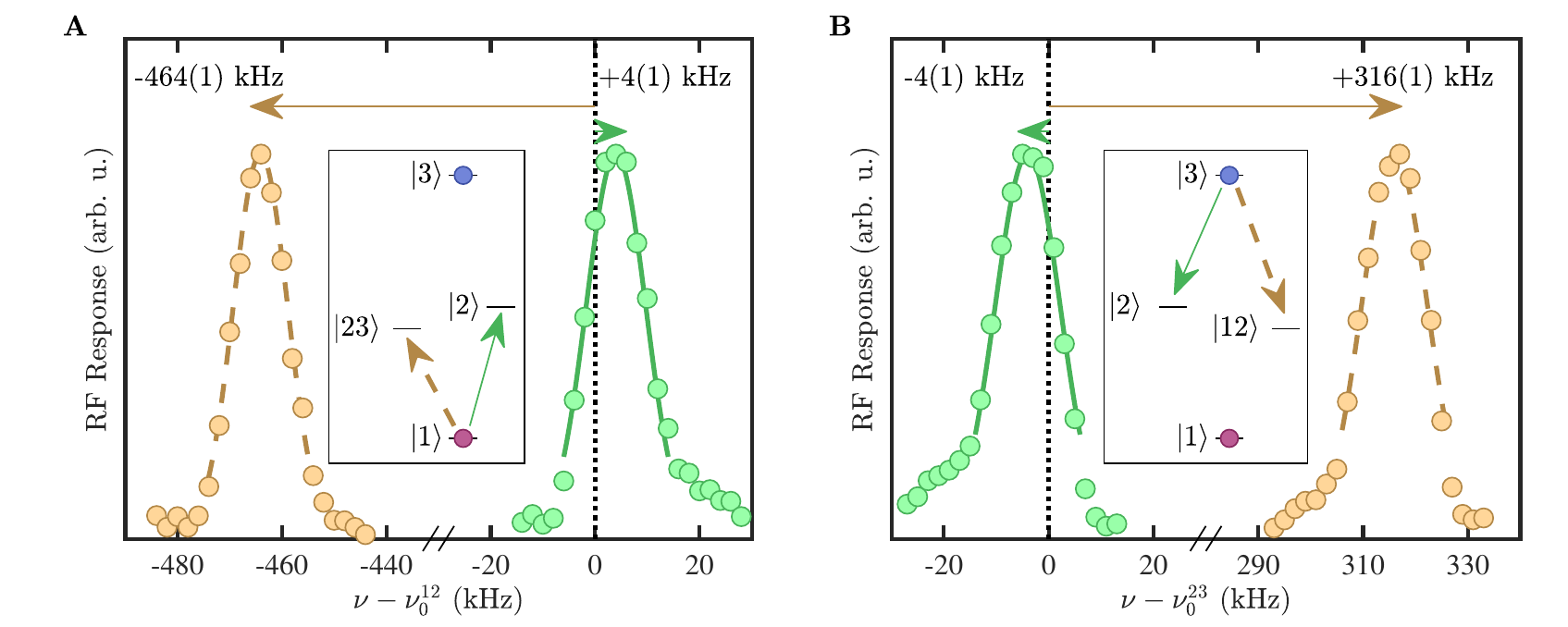}
\caption{Measurement of Feshbach dimer binding energies in a balanced $\ket{1}$-$\ket{3}$ mixture at $B=690~$G. \textbf{(A)} Spectroscopy on the transition $\ket{1}\rightarrow \ket{2}$. \textbf{(B)} Spectroscopy on the transition $\ket{3}\rightarrow \ket{2}$. For both measurements, the RF pulse time is 80 $\mu$s, with a transferred fraction of $\approx 25\%$ on the `free atom' peak (green data) and $\approx 10\%$ on the `bound' peak (orange data, vertically rescaled for comparison). The bare values $\nu_0^{12}=76.034$ MHz and $\nu_0^{23}=82.707$ MHz (vertical dotted lines) are calibrated on (noninteracting) fully polarized samples. The dashed and solid lines are gaussian fits from which peak frequencies are extracted. The cartoons are energy level diagrams with the transitions used (not to scale). }
\label{SFIG:dimerenergies}
\end{figure}

To evaluate this quantity, we performed RF spectroscopy of both Feshbach dimers $\ket{12}$ and $\ket{23}$ at $B=690~$G, see Fig.~\ref{SFIG:dimerenergies}. We find that the binding energies are respectively $\epsilon_\text{b}/h\approx320~\text{kHz}$ and $470~\text{kHz}$~\cite{footnoteDimerBindingEnergies}. Using the most deeply bound of the two dimers, $\log_{5/9}\left( 3 U_\text{box}/ \epsilon_\text{b} \right) + 1 \approx 4$ (where we have accounted for an additional loss from a final inelastic atom-dimer event).

\begin{figure}[H]
\centering
\includegraphics[width=0.9\columnwidth]{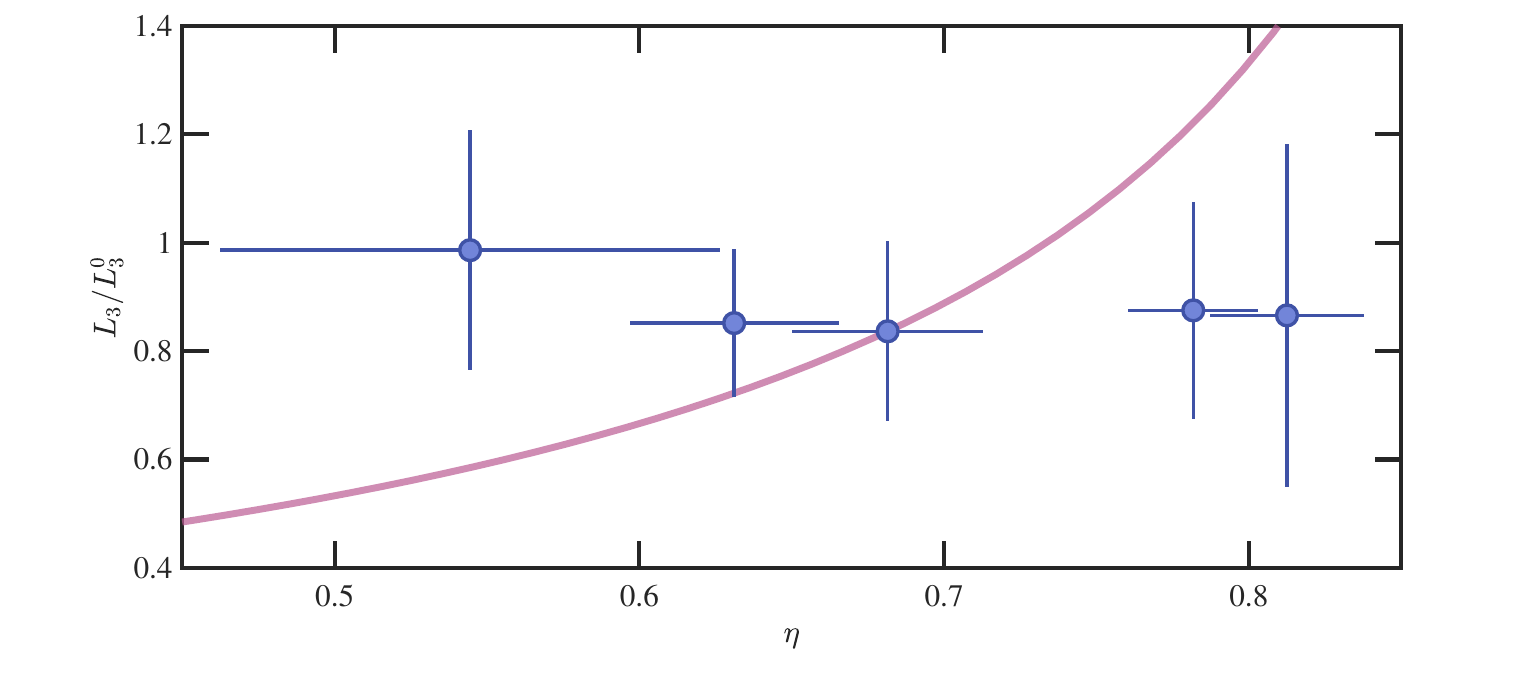}
\caption{Loss coefficient $L_3$ versus $\eta$. The data bins are the same as in Fig.~4E.  The prediction from the avalanche model is shown as the solid purple line (with an arbitrary scaling factor for comparison with the experimental data). Error bars are standard deviations within bins. }
\label{SFIG:k3andEta}
\end{figure}

Additionally, this model implies that the total loss rate would depend on the asymmetry. Indeed, for $\dot{n} = -3 L_3 n_1 n_2 n_3$, we would have within this model $3 L_3  =  K_3 (3 + c) \propto \left ( 1 + \frac{\eta}{1 - \eta} \right)$. Using the data binning of Fig.~4E, we plot $L_3$ vs $\eta$ in Fig.~\ref{SFIG:k3andEta}. The data displays no systematic variation of $L_3$ with $\eta$.

\section{VIII. Mean Free Path}

Avalanche processes require high collisional opacity of the product states in the medium.
However, we measure significant asymmetry even at low densities where the mean free path of typical recombination products $\ell$ is greater than the system size. 

For s-wave scattering of distinguishable particles, the collision cross section is ${\sigma = 4 \pi / (k^2 + a^{-2})}$, where $a$ is the relevant scattering length and $k = \sqrt{2 \mu E}/\hbar$ is the relative wavevector for a reduced mass $\mu$ and energy $E$. For a typical dimer $\ket{12}$ and free atom $\ket{3}$ formed after recombination, $k \approx 1\times10^7~\text{m}^{-1}$. 
The dimer (which scatters off all spin states) has the shorter mean free path:
\begin{align}
\ell \approx \left( n_1 \sigma_1 + n_2 \sigma_2 + n_3 \sigma_3 \right)^{-1} \sim 200~\mu \text{m}
\end{align}
where we have used $n_1 = n_2 = n_3 = 2\times10^{10}~\text{cm}^{-3}$ (corresponding to the lowest density bin of Fig.~4E of the main text), and $\sigma_j$ is the collisional cross section between the free atom $\ket{j}$ and the dimer $\ket{12}$. For this density, even a single secondary scattering is thus not very likely in a box of size $\sim150$ microns. Yet, even in those conditions we observe a large $\eta \gtrsim 0.5$. The result is similar for a typical $\ket{23}$ dimer and atom recombination product.

We repeated the experiment in a smaller, near-cubical box of size $\approx 30~\mu\text{m}$. In that box, a random particle would travel an average distance of $14~\mu$m (in the absence of collisions) before reaching the box boundary.  
Given an initial total density $n \approx 7\times 10^{11}~\text{cm}^{-3}$, $\ell > 21~\mu$m during the entire decay.  Even in that case, we find $\eta = 0.8(1)$, consistent with the measurements of the main text.  We conclude that an avalanche process is an unlikely explanation.

\section{IX. $L_3$ versus box depth}

\begin{figure}[H]
\centering
\includegraphics[width=0.9\columnwidth]{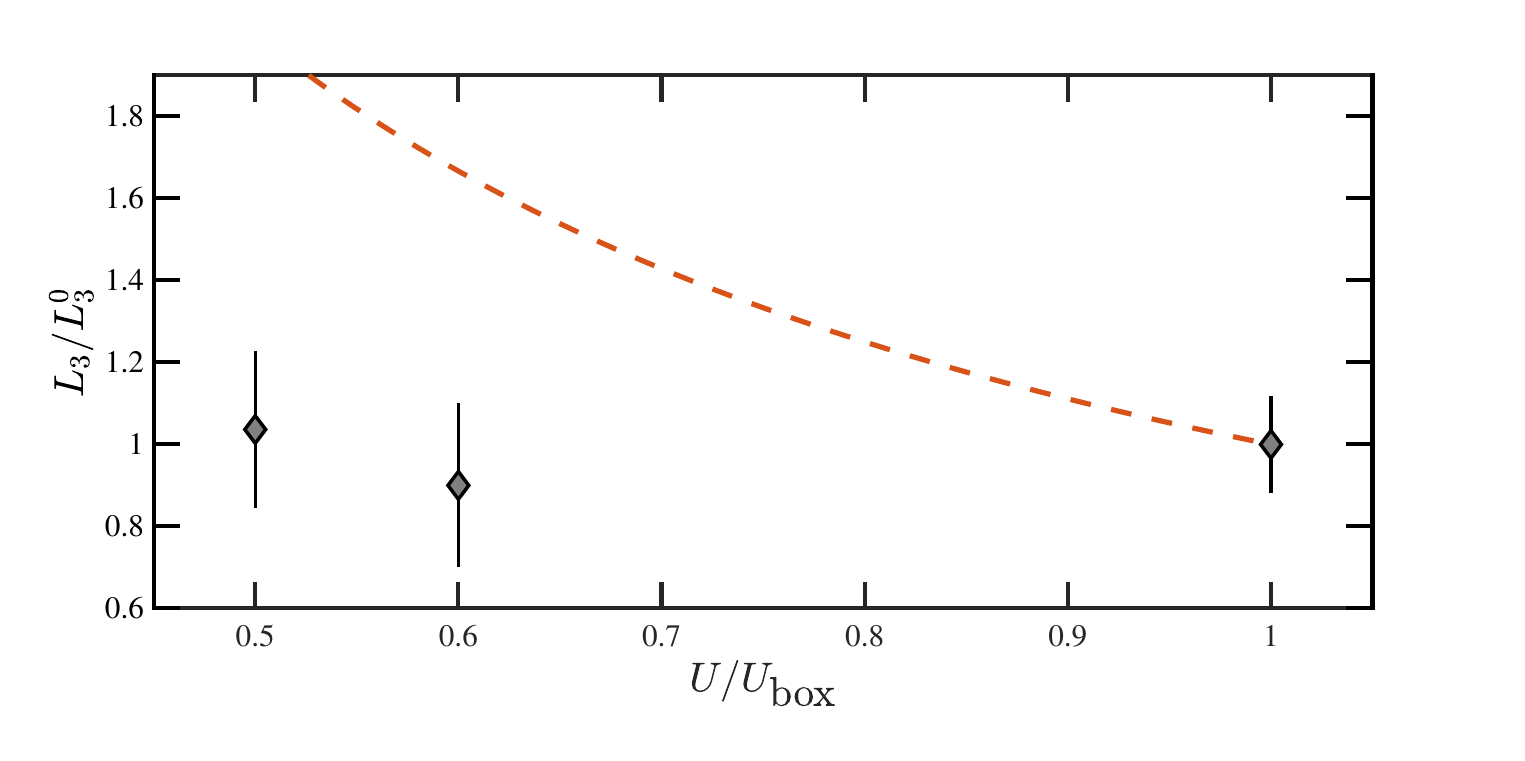}
\caption{$L_3$ versus trap depth $U$, where $U_\text{box} = k_\text{B} \times 1.6 ~\mu\text{K}$.  The red dashed line is the $1/U$ dependence expected for evaporation-dominated losses, rescaled by an arbitrary factor for comparison with the data.}
\label{SFIG:l3vsubox}
\end{figure}

Here, we test the influence of the trap depth on losses. We measure the decay of an unpolarized gas, lowering the box depth from its initial value $U_\text{box}$ to $U$ immediately after RF preparation of the three-component mixture. We see in Fig.~\ref{SFIG:l3vsubox} that the extracted $L_3$ shows no noticeable dependence on $U$ down to a depth of $U_\text{box}/2$. 
This indicates that evaporation plays essentially no role in the anomalous loss behavior. 

Suppose that inelastic losses deposit energy in the gas, which could have been released by recombination~\cite{weber2003three} or other mechanisms (e.g.~\cite{bouchoule2021losses}). This would lead to 
enhanced evaporation and thus a modification of the overall loss rate. 
For a polarized gas, these evaporative losses may be shared unequally among the spin states~\cite{parish2009evaporative} which could mimic the asymmetry we observe.  
We calculate the rate of losses assuming they are entirely due to evaporation and that the rate of deposited energy ($\epsilon_\text{b}$ per event) is balanced by the evaporation rate, such that
$ \dot{E} = \epsilon_\text{b} K_3 V n_1 n_2 n_3 + \alpha U \dot{N}=0$, where $U$ is the box depth, and $\alpha U$ is the average energy of a particle that leaves the trap ($\alpha \gtrsim 1$). This implies $\dot{n} =  -\frac{\epsilon_\text{b} K_3}{\alpha U} n_1 n_2 n_3$, and hence $L_3 \propto 1/U$, but we do not see evidence of this dependence (see dashed red line in Fig.~\ref{SFIG:l3vsubox}).

\begin{thebibliography}{72}%

\bibitem{vinen2002quantum}
W.~Vinen, J.~Niemela, Quantum turbulence, {\it Journal of Low Temperature
  Physics\/} {\bf 128}, 167--231 (2002).

\bibitem{pop2014coherent}
I.~M. Pop, {\it et~al.\/}, Coherent suppression of electromagnetic dissipation
  due to superconducting quasiparticles, {\it Nature\/} {\bf 508}, 369--372
  (2014).

\bibitem{thoennessen2004reaching}
M.~Thoennessen, Reaching the limits of nuclear stability, {\it Reports on
  Progress in Physics\/} {\bf 67}, 1187 (2004).

\bibitem{naidon2017Efimov}
P.~Naidon, S.~Endo, Efimov physics: a review, {\it Reports on Progress in
  Physics\/} {\bf 80}, 056001 (2017).

\bibitem{bottcher2020new}
F.~B{\"o}ttcher, {\it et~al.\/}, New states of matter with fine-tuned
  interactions: quantum droplets and dipolar supersolids, {\it Reports on
  Progress in Physics\/} {\bf 84}, 012403 (2020).

\bibitem{ueda2020quantum}
M.~Ueda, Quantum equilibration, thermalization and prethermalization in
  ultracold atoms, {\it Nature Reviews Physics\/} {\bf 2}, 669--681 (2020).

\bibitem{kagan1985effect}
Y.~Kagan, B.~Svistunov, G.~Shlyapnikov, Effect of {B}ose condensation on
  inelastic processes in gases, {\it JETP Lett\/} {\bf 42} (1985).

\bibitem{burt1997coherence}
E.~Burt, {\it et~al.\/}, Coherence, correlations, and collisions: What one
  learns about {B}ose-{E}instein condensates from their decay, {\it Physical
  Review Letters\/} {\bf 79}, 337 (1997).

\bibitem{laurent2017connecting}
S.~Laurent, {\it et~al.\/}, Connecting few-body inelastic decay to quantum
  correlations in a many-body system: A weakly coupled impurity in a resonant
  {F}ermi gas, {\it Physical Review Letters\/} {\bf 118}, 103403 (2017).

\bibitem{eigen2017universal}
C.~Eigen, {\it et~al.\/}, Universal scaling laws in the dynamics of a
  homogeneous unitary {B}ose gas, {\it Physical Review Letters\/} {\bf 119},
  250404 (2017).

\bibitem{he2020universal}
M.~He, C.~Lv, H.-Q. Lin, Q.~Zhou, Universal relations for ultracold reactive
  molecules, {\it Science Advances\/} {\bf 6}, eabd4699 (2020).

\bibitem{werner2022three}
F.~Werner, X.~Leyronas, Three-body contact for fermions. {I}. general relations,
  {\it arXiv:2211.09765\/}  (2022).

\bibitem{dieckmann2002decay}
K.~Dieckmann, {\it et~al.\/}, Decay of an ultracold fermionic lithium gas near
  a {F}eshbach resonance, {\it Physical Review Letters\/} {\bf 89}, 203201
  (2002).

\bibitem{o2002measurement}
K.~O’Hara, {\it et~al.\/}, Measurement of the zero crossing in a {F}eshbach
  resonance of fermionic {$^6$}{L}i, {\it Physical Review A\/} {\bf 66}, 041401
  (2002).

\bibitem{bourdel2003measurement}
T.~Bourdel, {\it et~al.\/}, Measurement of the interaction energy near a
  {F}eshbach resonance in a {$^6$}{L}i {F}ermi gas, {\it Physical Review
  Letters\/} {\bf 91}, 020402 (2003).

\bibitem{regal2003creation}
C.~A. Regal, C.~Ticknor, J.~L. Bohn, D.~S. Jin, Creation of ultracold molecules
  from a {F}ermi gas of atoms, {\it Nature\/} {\bf 424}, 47--50 (2003).

\bibitem{jochim2003pure}
S.~Jochim, {\it et~al.\/}, Pure gas of optically trapped molecules created from
  fermionic atoms, {\it Physical Review Letters\/} {\bf 91}, 240402 (2003).

\bibitem{randeria2012bcs}
M.~Randeria, W.~Zwerger, M.~Zwierlein, {\it The BCS-BEC Crossover and the
  Unitary {F}ermi Gas\/} (Springer, 2012), pp. 1--32.

\bibitem{modawi1997some}
A.~Modawi, A.~Leggett, Some properties of a spin-1 {F}ermi superfluid:
  Application to spin-polarized {$^6$}{Li}, {\it Journal of Low Temperature
  Physics\/} {\bf 109}, 625--639 (1997).

\bibitem{paananen2006pairing}
T.~Paananen, J.-P. Martikainen, P.~T{\"o}rm{\"a}, Pairing in a three-component
  {F}ermi gas, {\it Physical Review A\/} {\bf 73}, 053606 (2006).

\bibitem{schafer2007atomic}
T.~Sch{\"a}fer, What atomic liquids can teach us about quark liquids, {\it
  Progress of Theoretical Physics Supplement\/} {\bf 168}, 303--311 (2007).

\bibitem{o2011realizing}
K.~O'Hara, Realizing analogues of color superconductivity with ultracold alkali
  atoms, {\it New Journal of Physics\/} {\bf 13}, 065011 (2011).

\bibitem{adams2012strongly}
A.~Adams, L.~D. Carr, T.~Sch{\"a}fer, P.~Steinberg, J.~E. Thomas, Strongly
  correlated quantum fluids: ultracold quantum gases, quantum chromodynamic
  plasmas and holographic duality, {\it New Journal of Physics\/} {\bf 14},
  115009 (2012).

\bibitem{kurkcuoglu2018color}
D.~M. Kurkcuoglu, C.~S. de~Melo, Color superfluidity of neutral ultracold
  fermions in the presence of color-flip and color-orbit fields, {\it Physical
  Review A\/} {\bf 97}, 023632 (2018).

\bibitem{tajima2019quantum}
H.~Tajima, P.~Naidon, Quantum chromodynamics ({QCD})-like phase diagram with
  {E}fimov trimers and {C}ooper pairs in resonantly interacting {SU$(3)$}
  {F}ermi gases, {\it New Journal of Physics\/} {\bf 21}, 073051 (2019).

\bibitem{ottenstein2008collisional}
T.~B. Ottenstein, T.~Lompe, M.~Kohnen, A.~Wenz, S.~Jochim, Collisional
  stability of a three-component degenerate {F}ermi gas, {\it Physical Review
  Letters\/} {\bf 101}, 203202 (2008).

\bibitem{huckans2009three}
J.~Huckans, J.~Williams, E.~Hazlett, R.~Stites, K.~O’Hara, Three-body
  recombination in a three-state {F}ermi gas with widely tunable interactions,
  {\it Physical Review Letters\/} {\bf 102}, 165302 (2009).

\bibitem{williams2009evidence}
J.~Williams, {\it et~al.\/}, Evidence for an excited-state {E}fimov trimer in a
  three-component {F}ermi gas, {\it Physical Review Letters\/} {\bf 103},
  130404 (2009).

\bibitem{wenz2009universal}
A.~Wenz, {\it et~al.\/}, Universal trimer in a three-component {F}ermi gas,
  {\it Physical Review A\/} {\bf 80}, 040702 (2009).

\bibitem{lompe2010radio}
T.~Lompe, {\it et~al.\/}, Radio-frequency association of {E}fimov trimers, {\it
  Science\/} {\bf 330}, 940--944 (2010).

\bibitem{bause2021collisions}
R.~Bause, {\it et~al.\/}, Collisions of ultracold molecules in bright and dark
  optical dipole traps, {\it Physical Review Research\/} {\bf 3}, 033013
  (2021).

\bibitem{ji2022stability}
Y.~Ji, {\it et~al.\/}, Stability of the repulsive {F}ermi gas with contact
  interactions, {\it Physical Review Letters\/} {\bf 129}, 203402 (2022).

\bibitem{navon2021quantum}
N.~Navon, R.~P. Smith, Z.~Hadzibabic, Quantum gases in optical boxes, {\it
  Nature Physics\/} {\bf 17}, 1334--1341 (2021).

\bibitem{zurn2013precise}
G.~Z{\"u}rn, {\it et~al.\/}, Precise characterization of {$^6$}{Li} {F}eshbach
  resonances using trap-sideband-resolved {RF} spectroscopy of weakly bound
  molecules, {\it Physical Review Letters\/} {\bf 110}, 135301 (2013).

\bibitem{SuppMat}
See supplementary material.

\bibitem{footnoteMagneticLevitation}
For $B\geq 690$~G, the relative difference in magnetic moments between the
  spins is less than $h \times 10~\text{kHz}/\text{G}$ ($h$ is Planck's
  constant). Assuming that $\ket{1}$ is magnetically levitated against gravity
  ($|\text{d}B/\text{d}y| \approx 1~\text{G}/\text{cm}$), the residual forces on
  $\ket{2}$ and $\ket{3}$ are less than $1\%$ of gravity, an energy less than
  $k_\text{B} \times 5~\text{nK}$ over the size of our box, which is small
  compared to both our $E_\text{F}$ and $T$.

\bibitem{balancedexpt}
For some values of $B$, and depending on the imaging procedure, unpolarized
  samples can become polarized as a result of, for instance, two-component
  three-body recombinations in which the formed dimers are not
  imaged~\cite{ottenstein2008collisional}.

\bibitem{mehta2009general}
N.~P. Mehta, S.~T. Rittenhouse, J.~P.~D’Incao, J.~von~Stecher, C.~H. Greene,
  General theoretical description of {$N$}-body recombination, {\it Physical
  Review Letters\/} {\bf 103}, 153201 (2009).

\bibitem{petrov2003three}
D.~S. Petrov, Three-body problem in {F}ermi gases with short-range
  interparticle interaction, {\it Physical Review A\/} {\bf 67}, 010703 (2003).

\bibitem{nakajima2010nonuniversal}
S.~Nakajima, M.~Horikoshi, T.~Mukaiyama, P.~Naidon, M.~Ueda, Nonuniversal
  {E}fimov atom-dimer resonances in a three-component mixture of {$^6$}{Li},
  {\it Physical Review Letters\/} {\bf 105}, 023201 (2010).

\bibitem{lompe2010atom}
T.~Lompe, {\it et~al.\/}, Atom-dimer scattering in a three-component {F}ermi
  gas, {\it Physical Review Letters\/} {\bf 105}, 103201 (2010).

\bibitem{braaten2010efimov}
E.~Braaten, H.-W. Hammer, D.~Kang, L.~Platter, Efimov physics in {$^6$}{Li}
  atoms, {\it Physical Review A\/} {\bf 81}, 013605 (2010).

\bibitem{L3notationNote}
This is consistent with the notation of
  \cite{huckans2009three,williams2009evidence} and
  with~\cite{ottenstein2008collisional,braaten2010efimov} under the assumption
  $L_3 = K_3$.

\bibitem{L30comment}
Our measurement of $L_3^0$ at 690~G is similar to an unpolarized-gas
  measurement in a harmonic trap at a nearby field~\cite{huckans2009three}, but
  significantly smaller than a theoretical prediction~\cite{braaten2010efimov}.
  However, unitary saturation due to finite temperature effects could suppress
  the predicted resonant enhancement. Our measurements of $L_3$ above 800~G are
  similar to the values reported in~\cite{williams2009evidence}.

\bibitem{zwierlein2006fermionic}
M.~W. Zwierlein, A.~Schirotzek, C.~H. Schunck, W.~Ketterle, Fermionic
  superfluidity with imbalanced spin populations, {\it Science\/} {\bf 311},
  492--496 (2006).

\bibitem{partridge2006pairing}
G.~B. Partridge, W.~Li, R.~I. Kamar, Y.-a. Liao, R.~G. Hulet, Pairing and phase
  separation in a polarized {F}ermi gas, {\it Science\/} {\bf 311}, 503--505
  (2006).

\bibitem{nascimbene2009collective}
S.~Nascimbene, {\it et~al.\/}, Collective oscillations of an imbalanced {F}ermi
  gas: axial compression modes and polaron effective mass, {\it Physical Review
  Letters\/} {\bf 103}, 170402 (2009).

\bibitem{FelixYvan}
Y. Castin and F. Werner, private communication.

\bibitem{schuster2001avalanches}
J.~Schuster, {\it et~al.\/}, Avalanches in a {B}ose-{E}instein condensate, {\it
  Physical Review Letters\/} {\bf 87}, 170404 (2001).

\bibitem{zaccanti2009observation}
M.~Zaccanti, {\it et~al.\/}, Observation of an {E}fimov spectrum in an atomic
  system, {\it Nature Physics\/} {\bf 5}, 586--591 (2009).

\bibitem{langmack2013avalanche}
C.~Langmack, D.~H. Smith, E.~Braaten, Avalanche mechanism for the enhanced loss
  of ultracold atoms, {\it Physical Review A\/} {\bf 87}, 023620 (2013).

\bibitem{machtey2012universal}
O.~Machtey, D.~A. Kessler, L.~Khaykovich, Universal dimer in a collisionally
  opaque medium: experimental observables and {E}fimov resonances, {\it
  Physical Review Letters\/} {\bf 108}, 130403 (2012).

\bibitem{hu2014avalanche}
M.-G. Hu, R.~S. Bloom, D.~S. Jin, J.~M. Goldwin, Avalanche-mechanism loss at an
  atom-molecule {E}fimov resonance, {\it Physical Review A\/} {\bf 90}, 013619
  (2014).

\bibitem{zenesini2014resonant}
A.~Zenesini, {\it et~al.\/}, Resonant atom-dimer collisions in cesium: Testing
  universality at positive scattering lengths, {\it Physical Review A\/} {\bf
  90}, 022704 (2014).

\bibitem{parish2009evaporative}
M.~M. Parish, D.~A. Huse, Evaporative depolarization and spin transport in a
  unitary trapped {F}ermi gas, {\it Physical Review A\/} {\bf 80}, 063605
  (2009).

\bibitem{huang2020suppression}
C.-H. Huang, Y.~Takasu, Y.~Takahashi, M.~A. Cazalilla, Suppression and control
  of prethermalization in multicomponent {F}ermi gases following a quantum
  quench, {\it Physical Review A\/} {\bf 101}, 053620 (2020).

\end{thebibliography}

\begin{thebibliography}{3}%
\makeatletter
\let\auto@bib@innerbib\@empty
\makeatother
%</preamble>
\bibitem{schunck2008determination}
C.~H. Schunck, Y.~Shin, A.~Schirotzek, W.~Ketterle, Determination of the
  fermion pair size in a resonantly interacting superfluid, {\it Nature\/} {\bf
  454}, 739--743 (2008).

\bibitem{mukherjee2019spectral}
B.~Mukherjee, {\it et~al.\/}, Spectral response and contact of the unitary
  {F}ermi gas, {\it Physical Review Letters\/} {\bf 122}, 203402 (2019).

\bibitem{braaten2017lindblad}
E.~Braaten, H.-W. Hammer, G.~P. Lepage, Lindblad equation for the inelastic
  loss of ultracold atoms, {\it Physical Review A\/} {\bf 95}, 012708 (2017).

\bibitem{manzano2020short}
D.~Manzano, A short introduction to the {L}indblad master equation, {\it AIP
  Advances\/} {\bf 10}, 025106 (2020).

\bibitem{operatorDefNote}
In this section only, $N_j$ is defined as an operator. In the rest of the text
  it is an average value.

\bibitem{footnoteDimerBindingEnergies}
For both dimers, the measured binding energy is within $\sim 10\%$ of the
  universal expression $\epsilon_\text{b} = \hbar^2/(ma^2)$ and consistent with
  local measurements in a harmonically trapped gas at
  691~G~\cite{schunck2008determination}. Additionally, the shift of the free
  peak from the bare frequencies is consistent with the measurements of
  \cite{mukherjee2019spectral} for $T/T_\text{F}\approx 0.25$.

\bibitem{weber2003three}
T.~Weber, J.~Herbig, M.~Mark, H.-C. N{\"a}gerl, R.~Grimm, Three-body
  recombination at large scattering lengths in an ultracold atomic gas, {\it
  Physical Review Letters\/} {\bf 91}, 123201 (2003).

\bibitem{bouchoule2021losses}
I.~Bouchoule, L.~Dubois, L.-P. Barbier, Losses in interacting quantum gases:
  Ultraviolet divergence and its regularization, {\it Physical Review A\/} {\bf
  104}, L031304 (2021).

\end{thebibliography}
\end{document}